\documentclass[prd, aps, superscriptaddress, preprintnumbers, twocolumn,
floatfix, nofootinbib]{revtex4}

\usepackage{amsfonts} \usepackage{amsmath} \usepackage{amssymb}
\usepackage{bm} \usepackage{dcolumn} \usepackage{epsfig}
\usepackage{graphicx} \usepackage{graphics}
\usepackage[latin1]{inputenc} \usepackage{latexsym}
\usepackage{rotating} \usepackage{hyperref}

\usepackage{amsfonts} \usepackage{amsmath} \usepackage{amssymb}
\usepackage{bm} \usepackage{dcolumn} \usepackage{epsfig}
\usepackage{graphicx} \usepackage{graphics}
\usepackage[latin1]{inputenc} \usepackage{latexsym}
\usepackage{rotating} \usepackage{hyperref}

\newcommand\be{\begin{equation}} \newcommand\ba{\begin{eqnarray}}
\newcommand\ee{\end{equation}} \newcommand\ea{\end{eqnarray}}

\newcommand{\dd}{\mathrm d} \newcommand{\ex}{\mathrm e}
\newcommand{\GN}{G_{_\mathrm{N}}} \newcommand{\Hu}{\mathcal{H}}
\newcommand{\Ka}{\mathcal{K}}

\begin{document}

\title {Bouncing Cosmologies: Progress and Problems}

\author{Robert Brandenberger} \email{rhb@physics.mcgill.ca}
\affiliation{Physics Department, McGill University, Montreal, QC, H3A
2T8, Canada, and \\Institute for Theoretical Studies, ETH Z\"urich,
CH-8092 Z\"urich, Switzerland}

\author{Patrick Peter} \email{peter@iap.fr} \affiliation{Institut d'Astrophysique de Paris, UMR 7095,
and Institut Lagrange de Paris\\ 98 bis
boulevard  Arago, 75014 Paris, France \\ UPMC Universit\'e Paris 6 et CNRS,
  Sorbonne Universit\'es}

\date{\today}

\begin{abstract}

We review the status of bouncing cosmologies as alternatives to
cosmological inflation for providing a description of the very early
universe, and a source for the cosmological perturbations which are
observed today. We focus on the motivation for considering bouncing
cosmologies, the origin of fluctuations in these models, and the
challenges which various implementations face.

\end{abstract}

\pacs{98.80.Cq} \maketitle

\section{Motivation}

The inflationary scenario \cite{Guth} is the current paradigm of early
universe cosmology. Inflation solves several problems of Standard Big
Bang cosmology, and it gives rise to a causal theory of structure
formation \cite{Mukh} (see also \cite{Press}) which made a number of
predictions for cosmological observations which were subsequently
successfully verified. In spite of the phenomenological success,
inflation faces a  number of conceptual challenges (see e.g.
\cite{RHBrev} for a review of these problems) which motivate the
exploration of alternative early universe scenarios. Before mentioning
some of these challenges we must begin with a lightning review of
inflationary Universe cosmology.

\begin{figure}[t] \begin{center}
\includegraphics[scale=0.3]{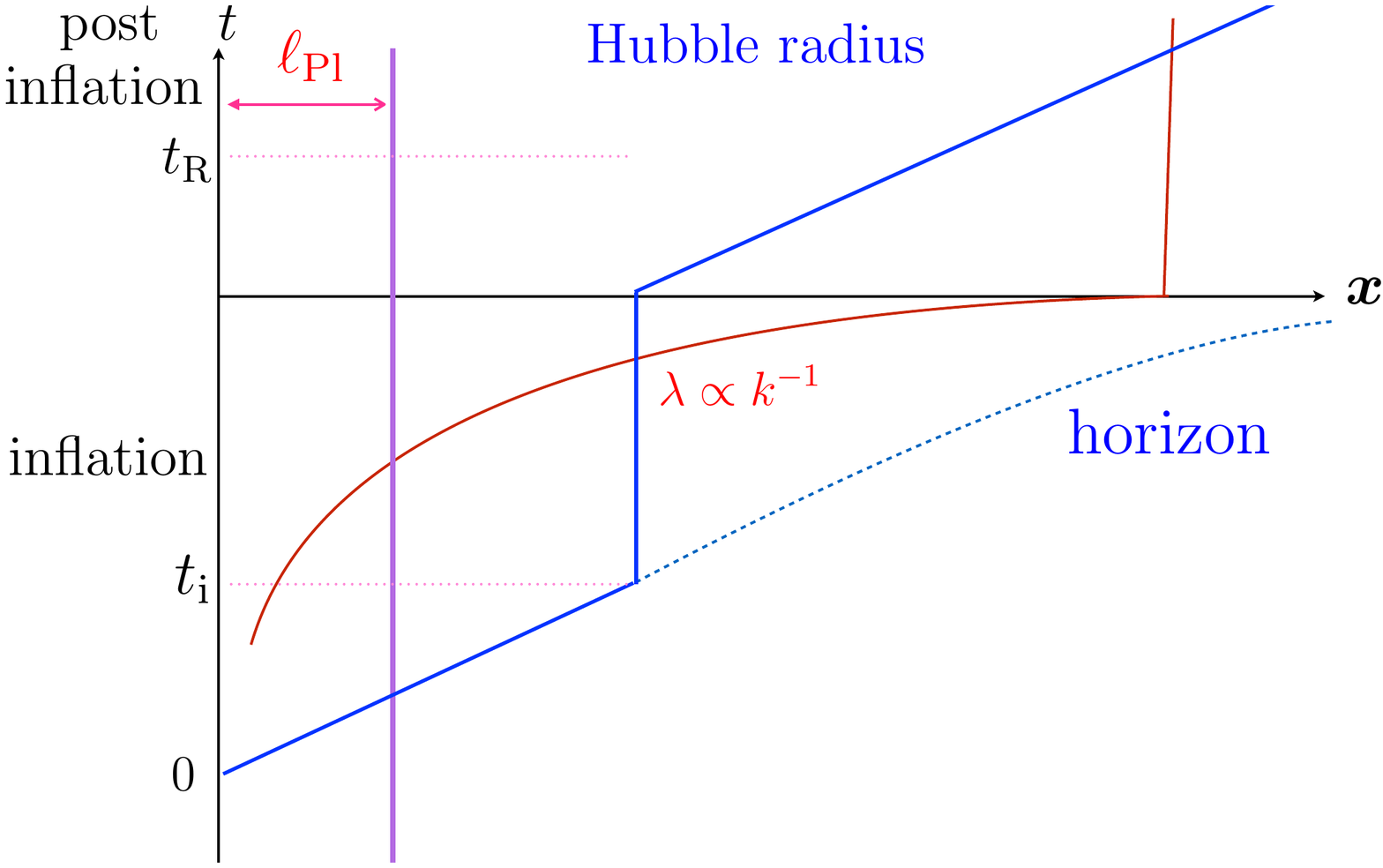} 
\caption{Space-time sketch of inflationary cosmology. The vertical axis
is time $t$, the period of accelerated expansion being between
$t_\mathrm{i}$ and $t_\mathrm{R}$ and the horizontal axis shows the
relevant physical distances. The horizon is represented by the
dotted curve and the Hubble radius by the full one. The
physical wavelength of a fluctuation mode is labeled by $\lambda$,
which is inversely proportional to the wave number $k$. The vertical
line shows the Planck length $\ell_\mathrm{Pl}$, emphasizing the
trans-Planckian zone of ignorance corresponding to length scales smaller
than $\ell_\mathrm{Pl}$. As is apparent, scales of cosmological interest
today originate in the zone of ignorance.} \label{FigScales}
\end{center} \end{figure}

According to the inflationary scenario, the universe underwent a period
of almost exponential expansion at some very early time. As a
consequence, the horizon expanded exponentially and became larger than
our past light cone - both evaluated at the time of recombination -
provided that the period of accelerated expansion was sufficiently long.
During this period, spatial curvature was also diluted. Any wavelength of
fluctuation was stretched quasi-exponentially during the period of
inflation so that the wavelengths corresponding to scales which are being
observed today in cosmological experiments were smaller than the Hubble
radius $H^{-1}(t)$ at the beginning of inflation, where $H(t)$ is the
expansion rate of space. The space-time geometry of inflationary
cosmology is sketched in Fig.~\ref{FigScales}. In this figure, the
vertical axis is time $t$, with $t = t_\mathrm{i}$ denoting the beginning
of the inflationary phase, and $t = t_\mathrm{R}$ the end; the horizontal
axis represents physical spatial distance. The Hubble radius is almost
constant between $t_\mathrm{i}$ and $t_\mathrm{R}$, and increases
linearly before $t_\mathrm{i}$ and after $t_\mathrm{R}$. The horizon is
shown as the dashed curve which equals the Hubble radius at the beginning
of the period of inflation but increases exponentially until
$t_\mathrm{R}$. The curve labeled $\lambda$ indicates the physical
wavelength of a cosmological fluctuation mode.

Assuming that the perturbations begin as quantum vacuum fluctuations
\cite{Mukh} deep inside the Hubble radius, one can compute the amplitude
of the resulting fluctuations today \cite{Mukh} (see \cite{MFB,
RHBpertrev} for reviews of the theory of cosmological perturbations). One
finds that the induced spectrum of fluctuations is approximately
scale-invariant and that the observed amplitude of fluctuations is
achieved if the Hubble expansion rate during the period of inflation was
of the order $H \sim 10^{13} {\rm GeV}$, which corresponds to an energy
density during the inflationary period which is of the order $\eta \sim
10^{16} {\rm GeV}$, the scale of particle physics ``Grand Unification''.
With this value of $H$, it turns out that the period of accelerated
expansion has to last for at least $50$ e-foldings in order for inflation
to be able to solve the horizon and flatness problems.

One key challenge for inflation is the singularity problem. If inflation
is realized by the dynamics of scalar matter fields coupled to Einstein
gravity, then the Hawking-Penrose singularity theorems \cite{Hawking} can
be extended \cite{Borde} to show that an inflationary universe is
geodesically past incomplete. Thus, there necessarily is a singularity
before the onset of inflation. Hence, the inflationary scenario cannot
yield the complete history of the very early universe. A bouncing
cosmological scenario naturally avoids this singularity problem, although
at the cost of having to introduce new physics to obtain the
bounce\footnote{We shall not here consider those mixed models in which a
contracting phase followed by a bounce leads to an inflationary era. Such
models enjoy the benefits of both paradigms, but also imply a higher
level of sophistication which, at the present time, may not be required
by the data; Occam's razor demands they should be introduced only at a
later stage if needed.}.

A second challenge which the inflationary scenario faces is the {\it
trans-Planckian} problem for fluctuations. If the period of accelerated
expansion lasts only slightly longer than the minimal amount of time
which it has to last in order to solve the problems of Standard Big Bang
cosmology, then the wavelengths of all scales of cosmological interest
today originate at sub-Planckian values at the beginning of inflation.
Hence, the origin and early evolution of cosmological fluctuations took
place in the trans-Planckian regime where General Relativity as the
description of space-time and quantum field theory as the description of
matter break down. As demonstrated in \cite{Jerome}, modifications of
the physics in the trans-Planckian region may lead to important
modifications of the predicted spectrum of cosmological perturbations.
In Fig.~\ref{FigScales}, we exhibit the trans-Planckian problem for
cosmological fluctuations: the shaded region for small distances
represents the ``trans-Planckian zone of ignorance''. The curve
labeled $\lambda$ indicates the wavelength of a cosmological fluctuation mode,
and shows that it emerges from the zone of ignorance.
Hence, in the absence of an embedding of inflation into a consistent
theory of quantum gravity, we cannot trust the standard computations in
inflation without implicitly assuming features of physics beyond the
Planck scale. There has been a lot of debate on how reasonable it is to
make assumptions which leave the standard computations unchanged (see
e.g. \cite{TPproblem}), or suggestions of how to impose initial
conditions on time-like ``new physics'' hypersurfaces which exclude the
trans-Planckian zone of ignorance \cite{NewPhysics}, leading to
oscillatory features in the predicted power spectra of fluctuations. For
a recent review of this problem, the reader is refered to
\cite{Jerome2}.

\begin{figure}[t] \begin{center}
\includegraphics[scale=0.3]{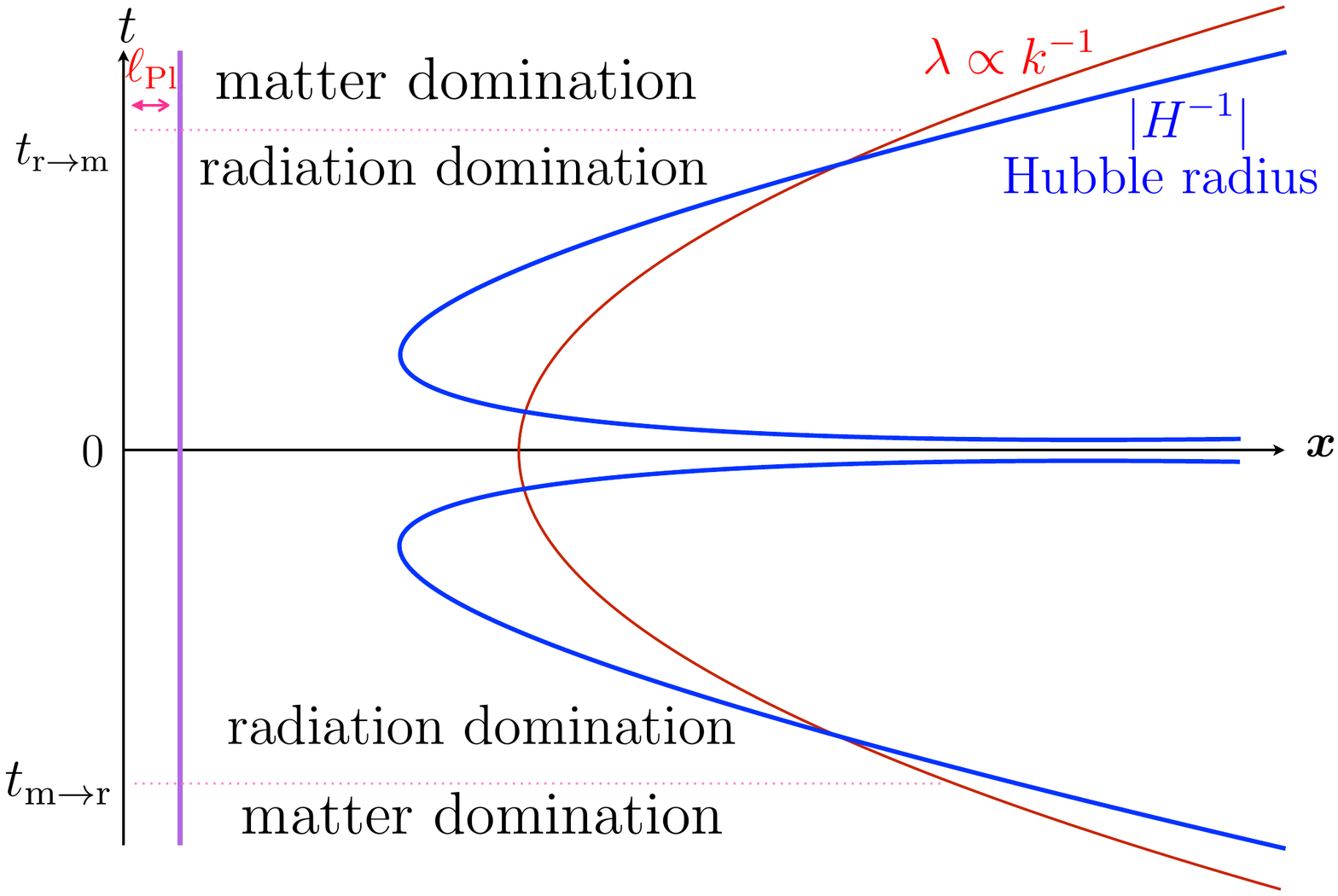} \caption{Space-time sketch
of a specific bouncing cosmology. The axes are as in Figure 1. The bounce
point corresponds to $t = 0$ at which point the Hubble radius
$H^{-1}(t)$, also shown, diverges ($H\to 0$) with the wavelength $\lambda
= a(t)/k$ of a typical mode depicted as well. The figure also shows
clearly why the trans-Planckian problem is not an issue in a bouncing
framework.}
\label{FigBounce} \end{center} \end{figure}

Bouncing cosmologies naturally avoid this trans-Planckian problem since
the length scale of fluctuations we observe remain many orders of
magnitude larger than the Planck length. To be specific, if the energy
scale of the bounce corresponds to the same energy scale as in typical
inflation models, then the wavelengths of scales coresponding to
observed cosmic microwave background (CMB) anisotropies were always
larger than $1$ mm. This is illustrated in the space-time sketch of
Fig.~\ref{FigBounce}. The vertical axis is time, with $t = 0$ being the
bounce time, the time when the spatial volume is minimal. This is thus
also the time when the wavelengths of cosmological perturbations are
minimal, and their values correspond to those in inflationary cosmology
at the end of the period of inflation, i.e. at the time $t_\mathrm{R}$.
The horizontal axis again represents physical spatial distance. What is
universal in bounce alternatives to inflation is that the evolution
after the bounce is the same as that in standard and inflationary
cosmology after the time $t_\mathrm{R}$. Different classes of bouncing
models predict different contracting phases. The example sketched in
Fig.~\ref{FigBounce} with a contracting phase which is the
time reverse of the standard cosmology phase of expansion
corresponds to a symmetric bounce as in the {\it
matter bounce} scenario \cite{Fabio, RHB_MBrev}.

As is evident from Fig.~\ref{FigBounce}, in bouncing cosmologies scales
of cosmological interest originate on sub-Hubble scales at early times
in the contracting phase, in the same way that they originate on
sub-Hubble scales early in the accelerating phase in inflationary
cosmology. Hence, it is possible to have a causal generation mechanism
for fluctuations, as in inflationary cosmology. The nature of this
structure formation mechanism depends on the specific bouncing model
being considered. The origin of the fluctuations 
is often (e.g. in the {\it Pre-Big-Bang}
\cite{PBB}, the Ekpyrotic \cite{Ekp} or matter bounce \cite{Fabio}
scenarios) also taken to be quantum vacuum perturbations, as in
inflation, but this is not always the case, {\it string gas cosmology}
\cite{BV} being an example in which the initial fluctuations are of
thermal nature \cite{NBV}.

A further conceptual problem of scalar field-driven inflation is the
basic sensitivity of the mechanism of obtaining exponential expansion to
the unknown solution of the cosmological constant problem. The
inflationary expansion of space is generated by the almost constant
potential energy density of the matter scalar field. However, quantum
matter has a field-independent vacuum energy which is much larger than
the upper bound on the current value of the cosmological constant, given
by the observed dark energy density (see e.g. \cite{JeromeReview}). There
must be some unknown mechanism which renders quantum vacuum energy
gravitationally inert. The challenge for inflationary cosmology is to
show that this mechanism does not also render the part of a scalar field
potential which has the equation of state of a cosmological constant
gravitationally inert (and thus eliminates the possibility of scalar
field-mediated inflation). For an interesting attack on this problem in
the context of the ``degravitation'' proposal \cite{Dvali} to address the
cosmological constant problem, see \cite{Patil}.

Alternatives to cosmological inflation do not solve the cosmological
constant problem. However, the dynamics of space-time expansion is then
{\sl a priori} not sensitive to what solves the problem (except if a
mechanism similar to that used for the inflationary phase is invoked for
the bouncing phase).

There is a completely different reason to be interested in alternatives
to inflation. The successful predictions which inflation has made for
the spectrum of cosmological perturbations are not specific to
inflation. In fact, a decade before the development of inflationary
cosmology, it was realized by Sunyaev and Zel'dovich \cite{SZ} and by
Peebles and Yu \cite{Peebles} that a roughly scale invariant spectrum of
adiabatic perturbations which at the time of equal matter and radiation
are present on super-Hubble scales will lead to the Sachs-Wolfe
scale-invariant large angle tail and to acoustic oscillations on degree
scales in the angular power spectrum of the CMB, it will lead to a power
spectrum of primordial density fluctuations which is scale invariant and has
superimposed small amplitude baryon acoustic oscillations, all features
which have now been observed. The inflationary scenario is the first
model which from first principles predicts such a primordial spectrum of
cosmological fluctuations, but it is not the only one. Therefore, even
if one believes that inflation is the correct paradigm for the early
universe, and that it gives the correct origin of cosmic structures, it
is important to work out alternative scenarios and to see in which way
these alternatives yield predictions (other than those mentioned above)
with which they can be differentiated from inflation. For example,
whereas simply observing gravitational waves on cosmological scales from
$B$-mode polarization is not at all a unique prediction of inflation (see
e.g. \cite{holy} for an elaboration on this point), measuring a slight
red tilt would be a rather distinctive prediction since it would allow
us to rule out some alternatives such as string gas cosmology which
predicts a slightly blue spectrum \cite{BNPV}. Working out distinctive
predictions of alternative models will provide new tests for inflation -
tests which, if passed will put inflation on a firmer footing, and if
failed will allow us to falsify the paradigm.

In the following, we will present three classes of bouncing cosmologies,
the {\it matter bounce} scenario \cite{Fabio}, models of {\it
Pre-Big-Bang} \cite{PBB} or Ekpyrotic \cite{Ekp} type, and {\it string
gas cosmology} \cite{BV, NBV}. In the next section we will briefly
introduce these scenarios and explain how to obtain a spectrum of
cosmological perturbations in agreement with current observations. New
physics, i.e. physics which goes beyond General Relativity and/or
standard matter theory (i.e. models of matter which obey the Null Energy
Condition) is required. In Section 3 we discuss some realizations of
cosmological bounces using extensions of gravity and/or the matter
sector. In Section 4 we illustrate some specific observational
signatures of these bouncing cosmology scenarios, signatures with which
they can be differentiated from the predictions of inflationary models.
We conclude with a discussion of some key challenges for bouncing
cosmologies.

Note that the goal of this article is to be pedagogical rather than
complete. We are not giving a complete list of alternatives to
inflation, nor are we discussing all bouncing scenarios (for a more
complete survey or older models see \cite{Novello}, and \cite{DBPP} for
a more recent review).

In the following we use the ``mostly negative'' convention for the
metric [see Eq.~\eqref{FLRW}]. Greek letters denote space-time indices
whereas Latin ones stand for spatial only indices.

\section{Origin of Scale-Invariance of Cosmological Fluctuations}

\subsection{Cosmological Perturbations}

In this section we will introduce four bouncing scenarios which have
been widely discussed in the literature, and we will show how to obtain
a scale-invariant spectrum of cosmological fluctuations at late times.
We must begin with a brief review of the basics of cosmological
perturbations. For more details the reader is referred to \cite{MFB,
RHBpertrev,PPJPU}.

The Friedman-Lema\^\i{}tre-Robertson-Walker (FLRW) metric of a
homogeneous and isotropic background space-time will be written as
\be \dd s^2  =  \dd t^2 - a^2(t) \dd\bm{x}^2  = a^2(t)\left( \dd\eta^2 -
\dd\bm{x}^2 \right), \label{FLRW} 
\ee 
where $t$ is the cosmic time, $\eta$ is conformal time, $\bm{x}$ are the
comoving spatial coordinates and $a(t)$ is the scale factor, in terms of
which the Hubble expansion rate is given by 
\be 
H(t) \equiv \frac{{\dot{a}}}{a} = \frac{1}{a} \left(
\frac{a'}{a}\right) \equiv \frac{\Hu}{a} \, , 
\ee 
an overdot and a prime denoting respectively derivatives with respect to
$t$ and $\eta$. In most cosmological models (string gas cosmology being
an exception), matter is modelled in terms of a scalar field $\varphi$
with a non-trivial background dynamics $\varphi_0(t)$.

Linear fluctuations of geometry and matter about the background can be
classified according to how they transform under spatial rotations.
There are scalar modes, vector modes and tensor modes (gravitational
waves). There are ten metric degrees of freedom plus the number of
degrees of freedom in the matter sector (one if the model contains only
a single scalar field like prototypical models of inflation). Four of
the metric degrees of freedom are scalar, four are vector, and the
remaining two are the two polarization modes of gravitational waves.
There are also four gauge modes which represent the invariance under
linearized space-time coordinate transformations. Two of the gauge modes
are scalar and two are vector. In the following we will not consider
vector modes since they decay in the expanding phase (however, they do
grow in the contracting phase \cite{Thorsten}).

We can choose a gauge in which the scalar metric fluctuations are
diagonal. In this gauge, the metric including linear fluctuations takes
the form 
\be \label{pertmetric} 
\dd s^2 = a^2 \left\{ \left(1 + 2 \Phi\right)
\dd\eta^2 - \left[\left(1 - 2\Psi\right) \delta_{ij} + h_{ij}\right] \dd
x^i \dd x^j \right\} , 
\ee 
where $\Phi(\bm{x}, t)$ and $\Psi(\bm{x}, t)$ denote the two scalar
metric fluctuations, and the transverse traceless tensor $h_{ij}$ (which
also depends on space and time) represents the gravitational waves. At
linear order, the equations for scalar metric perturbations and
gravitational waves decouple. We will first consider the scalar modes.

If the gravitational action is the Einstein-Hilbert action and if matter
contains no anisotropic stress, then it follows from the off-diagonal
spatial Einstein equations that the two scalar gravitational potentials
$\Phi$ and $\Psi$ coincide. The matter field can also be linearized
about its background: 
\be 
\varphi(\bm{x}, t) = \varphi_0(t) + \delta \varphi(\bm{x}, t) . 
\ee

The Einstein constraint equations relate the two scalar fluctuations
$\Phi$ and $\delta \varphi$. Physically this is easy to understand: a
matter perturbation has a gravitational effect on the metric and leads
to a metric perturbation. As a consequence of this contraint there is a
single scalar metric degree of freedom.

The equations of motion for cosmological fluctuations can be obtained by
linearizing the full Einstein equations about the background metric.
More easily, the equations can be obtained by expanding the action of
matter plus gravity to second order about the background action. Since
the background satisfies the equations of motion, terms linear in
cosmological fluctuations cancel out in the action, leaving the
quadratic terms as the leading fluctuation terms. Working in terms of
the action allows the identification of the variable $v({\bm{x}}, t)$ in
terms of which the action has canonical kinetic terms \cite{Sasaki,
Mukh2}. This variable is 
\be 
v = a \left( \delta \varphi + \frac{z}{a} \Phi \right) , 
\ee 
where 
\be 
z \, = \, \frac{a \varphi_0^{\prime}}{{\cal H}} \, . 
\ee 
It turns out that the canonical
variable $v$ has an important physical meaning: it is proportional to
the curvature fluctuation $\zeta$ in comoving gauge (gauge in which
$\delta \varphi = 0$): 
\be 
\zeta \equiv \frac{v}{z}\label{zeta} . 
\ee 
It is this variable which determines the late time curvature
fluctuations. Hence, in every early universe model, we need to be able to
compute the power spectrum of $\zeta$ which is defined via 
\be 
P_{\zeta}(k) \, = \, k^3 |\zeta(k)|^2 \, , 
\ee 
with $\zeta(k)$ the Fourier transform of $\zeta$ \footnote{Here we
are using a convention for the Fourier transform in which the
Fourier modes have mass dimension $-3/2$ plus the mass dimension
of the position space quantity. }.
Conventionally, one introduces a scalar index $n_\mathrm{s}$ to describe
the slope of the spectrum, namely 
\be 
P(k) \sim  k^{n_\mathrm{s} - 1}. 
\ee

The canonical variable $v$ evolves like a scalar field with a
time-dependent mass, the time dependence being set by the dynamics of
the cosmological background 
\be 
v^{\prime \prime} - \nabla^2 v - \frac{z^{\prime \prime}}{z} v \, =
\, 0 \, , 
\ee 
where the $\nabla$ operator corresponds to the gradient operator with
respect to the comoving coordinates ${\bm{x}}$. At the linear level, we
can expand $v({\bm{x}}, t)$ into spatial Fourier modes which all evolve
independently. In Fourier space, the equation of motion becomes 
\be 
\label{FourierEoM} v_k^{\prime \prime} + \left( k^2 -
\frac{z^{\prime \prime}}{z} \right) v_k \, = \, 0 \, , 
\ee 
where $k$ labels the comoving wave vectors. For vacuum initial
conditions imposed at some time $t_\mathrm{i}$, the initial value of $v_k$ is
given by 
\be
 \label{IC} v_k(t_\mathrm{i}) \, = \, \frac{1}{\sqrt{2 k}} \, .
 \ee 
Note that the corresponding power spectrum is blue, i.e. there is more
power on short wavelengths. The value of the index $n_\mathrm{s} = 1$
corresponds to a {\it scale-invariant} spectrum, i.e. a power spectrum
which is independent of scale $k$.

If the equation of state (i.e. the ratio between pressure and energy
density) of the cosmological background is constant in time, then
$z(\eta) \sim a(\eta)$ and it then follows that $z^{\prime \prime} / z$
is equal (up to a constant of order 1) to ${\cal H}^2$. Thus, from Eq.
(\ref{FourierEoM}) it follows that the ``squeezing term'' $z^{\prime
\prime} / z$ is negligible compared to the $k^2$ term for wavelengths
smaller than the Hubble radius. This implies that the variable $v$ will
undergo oscillations with constant amplitude on sub-Hubble scales. On
super-Hubble scales it is the $k^2$ term which is negligible compared to
the squeezing term. Hence, we conclude that as a length scale crosses
the Hubble radius (e.g. during the inflationary phase), the fluctuations
will stop to oscillate (they ``freeze out''), and the amplitude will
begin to increase as a consequence of the squeezing term (the
fluctuations are ``squeezed''). The equation of motion for $v_k$ has two
fundamental solutions. In the case of an expanding background, the
dominant solution on super-Hubble scales is given by 
\be 
v_k(\eta) \, \sim \, z( \eta) \, . 
\ee

Let us first apply this formalism and show that a period of exponential
expansion leads to a scale-invariant spectrum of cosmological
perturbations. The time $t_H(k)$ when the mode with wavenumber $k$
crosses the Hubble radius is given by 
\be \label{crossing} 
a^{-1}\left[ t_H\left( k \right) \right] k \, = \, H \, . 
\ee 
Before $t_H(k)$, the amplitude of $v_k$ is constant, afterwards it
increases in proportion to $z$. Hence, the power spectrum of $\zeta$ at
some late time $t$ is given 
\begin{equation}
\begin{split}
P_{\zeta}(k, \eta) =&  \, k^3 z^{-2}(\eta) |v_k(\eta)|^2  \\
\simeq& \, k^3 z^{-2}(\eta) \left\{
\frac{z(\eta)}{z\left[\eta_H(k)\right]} \right\}^2 |v_k(t_\mathrm{i})|^2
\\ \simeq& \, \frac{1}{2} \left\{
\frac{a\left[t_H(k)\right]}{z\left[t_H(k)\right]} \right\}^2 H^2
\end{split}
\end{equation}
where in the final step we have used (\ref{IC}) and (\ref{crossing}).
The $k$-dependence has cancelled out and we have a scale-invariant power
spectrum. Long wavelength modes are squeezed more than short wavelength
ones. It is this effect which leads to the conversion of an initial
vacuum spectrum into a scale-invariant one.

Exponential expansion is not the only expansion history which is able to
convert a vacuum spectrum into a scale-invariant one. As we will see in
the following subsection, a matter-dominated contraction phase leads to
exactly the same preferential squeezing of long wavelength modes,
compared to short wavelength ones, and which is able to convert a vacuum
spectrum into a scale-invariant one \cite{Wands}.

In the case of more than one matter field there are extra scalar degrees
of freedom which are usually called {\it entropy modes}. Entropy
fluctuations induce a growing curvature mode. Let us consider the case
when a scalar field $\varphi$ dominates the background, and a second
scalar field $\chi$ is the entropy mode with energy density fluctuation
$\delta \rho_{\chi}$. In this case, the induced curvature mode is given
by the following equation (see e.g. \cite{Gordon, Malik}) 
\be 
\label{conversion} {\dot{\zeta}} \, = \,  H \left( c^2_{\varphi} -
c^2_{\chi} \right) \frac{\delta \rho_{\chi}}{\rho_{\varphi} +
p_{\varphi}} \, , 
\ee 
where $\rho_{\varphi}$ and $p_{\varphi}$ are energy density and
pressure in the $\varphi$ field, $\delta \rho_{\chi}$ is the energy
density fluctuation of the $\chi$ field, and $c_\varphi$ and $c_\chi$ 
are the speeds of sound of the $\varphi$ and $\chi$ fluids, respectively.

In cosmological models in which the adiabatic mode (the fluctuations in
the dominant matter component) has a blue spectrum (as in the
Pre-Big-Bang and Ekpyrotic scenarios), it is possible to obtain a
scale-invariant spectrum of curvature fluctuations at late times via the
conversion equation (\ref{conversion}) if the entropy fluctuations
acquires a scale-invariant spectrum.

To conclude this subsection let us turn to an analysis of gravitational
waves. The transverse traceless tensor $h_{ij}$ for the tensor modes
[see Eq.~\eqref{pertmetric}] can be expanded in terms of the two polarization
states 
\be \label{gravwave} 
h_{ij}(\bm{x}, \eta) \, = \, h_{+}(\bm{x},
\eta)\epsilon^{+}_{ij} + h_\times(\bm{x}, \eta)\epsilon^{\times}_{ij} \, , 
\ee 
where $\epsilon^{+}_{ij}$ and $\epsilon^{\times}_{ij}$ are two fixed
polarization tensors, and $h^{+}$ and $h^{\times}$ are the amplitude
functions for these modes. Each of these two modes evolves independently
at linear order.

Inserting the ansatz (\ref{gravwave}) for an individual polarization
state into the Einstein action, we find that the variable in terms of
which the perturbed action has canonical kinetic term is 
\be 
\mu(\bm{x}, \eta) \, \equiv \, a(\eta) h(\bm{x}, \eta) \, , 
\ee 
resulting in the following equation of motion for each Fourier mode 
\be \label{tensorEoM} 
\mu_k^{\prime \prime} + \left( k^2 -
\frac{a^{\prime \prime}}{a} \right) \mu_k \, = \, 0 \, . 
\ee

Comparing the equations of scalar and tensor modes, we see that they are
very similar except that the squeezing function $z^{\prime \prime} / z$
is replaced by $a^{\prime \prime}/a$ in the case of gravitational waves.
Both gravitational waves and scalar fluctuations oscillate on sub-Hubble
scales, both freeze out at Hubble radius crossing (more precisely when
$k^2$ becomes equal to the squeezing function), and both are squeezed on
super-Hubble scales. If the equation of state of matter is independent
of time, then $z \propto a$ and the squeezing functions are identical.

\subsection{Matter Bounce Scenario}

The ``matter bounce'' scenario \cite{Fabio} is based on the duality
\cite{Wands} (see also \cite{Edna4}) between the evolution of the canonical fluctuation
variables in an exponentially expanding period and in a
contracting phase with pressure $p = 0$. In an expanding universe, the growing
mode of $v_k$ is the mode which is proportional to $z(\eta)$ and the
second mode is decaying. In contrast, in a contracting universe the role
of the modes is exchanged: the mode proportional to $z$ is decaying, and
it is the second mode which is growing. In a matter dominated phase of
contraction we have $a(\eta) \sim \eta^2$, and the solution of the mode
equation (\ref{FourierEoM}) on super-Hubble scales is 
\be 
v_k(\eta) \, = \, c_1 \eta^2 + c_2 \eta^{-1} \, , 
\ee 
where the constants $c_1$ and $c_2$ are set by the initial conditions.
The second mode is the growing one. It is the fact that the equation of
state is that of cold matter ($p = 0$) which leads to the characteristic
scaling $v_k \sim \eta^{-1}$ which, as we show below, leads to the
conversion of a vacuum spectrum to a scale-invariant one on super-Hubble
scales.

Given the scaling $v_k \sim \eta^{-1}$, it is easy to show that an
initial vacuum spectrum on sub-Hubble scales early in the contracting
phase is converted into a scale-invariant one. The calculation follows
the one done in the previous subsection in the case of an exponentially
expanding background: 
\begin{equation}
\begin{split}
P_{\zeta}(k, \eta) \, =& \, k^3 z^{-2}(\eta)
|v_k(\eta)|^2  \\ \simeq& \, k^3 z^{-2}(\eta) \left[
\frac{\eta_H(k)}{\eta} \right]^2 |v_k(t_\mathrm{i})|^2 \, . 
\end{split}
\label{matterpower}
\end{equation}
Since ${\cal{H}} \sim \eta^{-1}$ and since Hubble radius crossing is
given by $k^2 = {\cal{H}}^2$, we find that 
\be 
\eta_H(k) \, \sim \, k^{-1} \, . 
\ee 
Inserting this result into (\ref{matterpower}) and making use of vacuum
initial conditions leads to a scale-invariant spectrum of perturbations
on super-Hubble scales, as in the case of inflationary cosmology.

In the case of scalar field-driven inflation, the spectrum of
cosmological perturbations is not exactly scale-invariant, but it has a
slight red tilt. The same is true in the matter bounce scenario: if we
add a component to matter which corresponds to the current dark energy
(e.g. a small cosmological constant), then a slight red tilt results
\cite{Ewing} (see also \cite{Cai:2015vzv}).

Since in the matter bounce scenario the universe begins large and cold,
it is reasonable to consider vacuum initial fluctuations. It is, however, also
possible to consider thermal initial fluctuations. In this case, the
resulting spectrum of curvature fluctuations after the bounce is
not scale-invariant. There is, however, a particular equation of state
in a contracting phase \cite{thermal} which allows one to transform
a thermal particle spectrum into a scale-invariant one.

\subsection{Pre-Big-Bang Scenario}

The Pre-Big-Bang scenario \cite{PBB} (see \cite{PBBreview} for an
in-depth review) is an approach to superstring cosmology. If superstring
theory is indeed the correct theory which unifies all four forces of
nature and provides a quantum theory of gravity, then we should expect
that stringy effects will be important in the very early universe. One
aspect of string theory is that the graviton is not the only massless
mode. In addition, there is a dilaton and an antisymmetric tensor field.
It is usually assumed that the dilaton is fixed at late times, but in
the very early universe we should expect it to be dynamical.

Neglecting, for the moment, the antisymmetric tensor field, the massless
sector of string theory to which the graviton belongs is given by
dilaton gravity. In the string frame the action is 
\be 
S \, = \, \int \dd^{d + 1}x \sqrt{-g} \ex^{- \phi} \left( R + g^{\mu
\nu} \partial_{\mu} \phi \partial_{\nu} \phi \right) \, , 
\ee 
where $R$ is the Ricci scalar of the string frame metric, $g$ is its
determinant, $\phi$ is the string frame dilaton, and $d$ is the number
of spatial dimensions.

A conformal transformation takes the action into the Einstein frame. In
the case of a homogeneous and isotropic space-time metric, this
transformation takes the form 
\begin{equation}
\begin{split}
{\tilde a} \, &= a \ex^{- \phi /(d -1)}  \\
{\tilde \phi} \, &= \phi
\sqrt{\frac{2}{d - 1}} , 
\end{split}
\end{equation}
where 
tilde quantities are in the Einstein frame. The Einstein frame action is 
\be 
S \, = \, \int \dd^{d + 1}x \sqrt{- {\tilde g}} \left( {\tilde R} -
\frac{1}{2} {\tilde g}^{\mu \nu} \partial_{\mu} {\tilde \phi}
\partial_{\nu} {\tilde \phi} \right)\, .
 \ee

The dilaton gravity action has time reflection symmetry, but it also has
a {\it scale factor duality symmetry} which is related to the T-duality
symmetry of string theory discussed in more detail in the
subsection on string gas cosmology. In the case of a homogeneous and
isotropic metric, this transformation is 
\ba 
a \, &\rightarrow& \, a^{-1} \\ {\bar \phi} \, &\rightarrow& \,
{\bar \phi} \, , \nonumber 
\ea 
where ${\bar \phi}$ is the ``invariant'' dilaton defined via 
\be 
{\bar \phi} \, = \, \phi - 2 d\, {\rm ln} a \, . 
\ee

If we minimally couple perfect fluid matter to space-time in the
Einstein frame, then in the case of $d = 3$ and an equation of state of
radiation, we obtain the usual expanding radiation-dominated solution 
\ba a(t) \, &\sim& \, t^{1/2} \\ \phi(t) \, &=& \, {\rm const} \, .
\nonumber \ea 
This is the post-big-bang expanding solution. If we first perform a time
reflection transformation followed by a duality transformation, we obtain
a super-exponentially expanding solution with 
\ba 
a(t) \, &\sim& \, (-t)^{-1/2} \\ \phi(t) \, &\sim& - 3 \ln (-t)
\, \nonumber 
\ea 
which has growing string coupling constant and corresponds to matter
with a stringy equation of state $p = - \frac13 \rho$. This is the so-called
``Pre-Big-Bang'' branch. In the Einstein frame, the Pre-Big-Bang solution
is a contracting one. In both the Einstein and string frames, it is easy
to verify that constant comoving scales exit the Hubble radius in the
Pre-Big-Bang phase, as they do in the matter bounce scenario.

Note that both of the above solutions are singular at $t = 0$. The idea
of the Pre-Big-Bang scenario is that new physics takes over near $t = 0$
and smoothly connects the Pre-Big-Bang and post-big-bang branches. In
the Einstein frame, this thus leads to a bouncing cosmology.

Also for other equations of state of matter we can find a Pre-Big-Bang
branch of an expanding solution with decreasing curvature. In the case
of pure dilaton gravity, the contracting branch is given by the Einstein
frame metric scaling as \cite{PBBreview} 
\be 
a(\eta) \, \sim \, \eta^{1/2} \, , 
\ee 
where $\eta$ is approaching $0$ from negative values. In this case, the
mode equation (\ref{FourierEoM}) for scalar cosmological fluctuations
has super-Hubble solutions 
\be \label{sol1} 
v_k(\eta) \, \sim \, |\eta|^{1/2} 
\ee 
and 
\be \label{sol2} 
v_k(\eta) \, \sim \, |\eta|^{1/2} \ln (k |\eta|) \, .
\ee 
The second solution is the dominant one as $\eta \rightarrow 0$.

From (\ref{sol1}) and (\ref{sol2}), it is easy to see that in contrast to
the matter bounce case, fluctuations are damped on super-Hubble scales.
Hence, an initial vacuum spectrum is not tilted towards the red, but
towards the blue. The power spectrum is (up to logarithmic corrections)
\be 
P_k(v) \, \sim \, k^3 |v_k\left[\eta_H(k)\right]|^2
\frac{\eta}{\eta_H(k)} 
\ee 
Assuming an initial vacuum spectrum, the second term on the right hand
side scales as $k^{-1}$, but since $\eta_H(k) \sim k^{-1}$ the third
term scales as $k$, and hence we get an $n_\mathrm{s} = 4$ spectrum 
\be 
P_k(v) \, \sim \, k^3 \, . 
\ee 
The tensor modes acquire a spectrum with the same slope.

 In order to obtain a scale-invariant spectrum of fluctuations in the
 Pre-Big-Bang scenario, we need to either dramatically deform the
 background or else invoke the entropy mechanism given by
 (\ref{conversion}). In the Pre-Big-Bang setup, there is a natural
 candidate for the entropy field $\chi$ of (\ref{conversion}): the axion
 related to the antisymmetric tensor field $H$ appearing in the massless
 spectrum of string theory. A careful analysis \cite{Copeland} shows
 that, in the presence of dynamical extra dimension, it is possible to
 obtain a scale-invariant spectrum of axion energy density fluctuations
 which, via (\ref{conversion}), will seed a scale-invariant spectrum of
 curvature fluctuations.

\subsection{Ekpyrotic Scenario}

The Ekpyrotic scenario \cite{Ekp} is also motivated by superstring
theory, and more specifically by Horava-Witten theory \cite{HW}, a
specific realization of M-theory in which space-time is 11 dimensional.
Six spatial dimensions are compactified with a fixed radius, but there
is one other spatial dimensions which is orbifolded, i.e. it has the
structure $S_1 / {\cal Z}_2$. At the orbifold fixed points there are
boundary branes, one of them being our space-time, the other containing
hidden sector fields which only couple gravitationally to our world.
This model was introduced for reasons of superstring phenomenology, and
the mass scale of the orbifold direction must be of the order of the
scale of particle physics Grand Unification.

The authors of \cite{Ekp} add a negative exponential potential which
causes the two boundary branes to approach each other. In the Einstein
frame, the period when the two branes are approaching each other is a
contracting phase. The time when the two branes collide corresponds to a
singular bounce from the point of view of the low energy effective field
theory.

The low energy effective field theory of the Ekpyrotic scenario is given
by Einstein gravity minimally coupled to a scalar field $\phi$ with a
negative exponential potential 
\be 
V(\phi) \, = -| V_0 | \exp\left( \sqrt{\frac{2}{p}}
\frac{\phi}{m_\mathrm{Pl}} \right)\label{PotEkp} 
\ee 
with index $p \ll 1$, where $m_\mathrm{Pl}$ is the Planck mass. The
field $\phi$ is related to the separation of the branes via $\phi \sim
\ln r$. The key feature of the Ekpyrotic scenario is that the
equation of state parameter $w$ is much larger than $1$ during the
contracting phase. This is a consequence of the potential being
a negative exponential. This results in the potential energy partially canceling the
kinetic energy, whereas both kinetic and potential terms yield positive
contributions to the pressure.

An equation of state $w \gg 1$ corresponds to very slow contraction
since 
\be 
(1 + w) \, = \, \frac{2}{3 \beta} 
\ee 
if the scale factor $a(t)$ scales as 
\be 
a(t) \, \sim \, (-t)^{\beta} \, . \label{atp}
\ee 
In conformal time we have 
\be 
a(\eta) \, \sim \, (- \eta)^{\alpha} \label{aetap} 
\ee 
with 
\be 
\alpha \, = \, \frac{\beta}{1 - \beta} \, . 
\ee 
Hence, as $w \gg 1$ we have $\alpha \sim 0$. Returning to the mode
equation (\ref{FourierEoM}) -- and to the corresponding equation for
gravitational waves \eqref{tensorEoM} --, we see that the spectrum of $v$
is not changed on super-Hubble scales. Thus, if we start with vacuum
initial conditions, then the spectrum of both curvature fluctuations and
gravitational waves remains a vacuum spectrum 
\be 
P_k(v) \, \sim \, k^2 \, , \label{nt3}
\ee 
i.e. a $n_\mathrm{s} = 3$ spectrum.

As realized in \cite{NewEkp}, a spectator scalar field $\chi$ evolving
in a negative exponential potential acquires fluctuations with a
scale-invariant spectrum. This can easily be seen in the absence of
gravity, where the equation of motion for the linearized fluctuation
modes becomes 
\be 
{\ddot \delta \chi} + \left( k^2 - \frac{2}{t^2} \right) \delta \chi
\, = \, 0 \, . 
\ee 
This is the same equation which the curvature fluctuations in
inflationary cosmology obey (in conformal time), and hence the same
analysis which was done in the case of inflationary fluctuations shows
that initial vacuum perturbations on sub-Hubble scales acquire a
scale-invariant spectrum on super-Hubble scales.

Thus, one way to obtain a scale-invariant spectrum of curvature
perturbations in the Ekpyrotic scenario is to posit the existence of a
second scalar field $\chi$ which evolves in a similar negative
exponential potential as the field $\phi$ which generates the Ekpyrotic
contraction. In order for scale-invariant $\chi$ field fluctuations (as
opposed to $\chi$ density fluctuations) to lead to a scale-invariant
spectrum of curvature fluctuations, a coupling between the two fields
$\phi$ and $\chi$ at the background level is required (see e.g.
\cite{Gordon}).

There is, however, another way to generate a scale-invariant spectrum of
curvature fluctuations at late times in the expanding phase. Note that,
up to now, we have been speaking about the curvature fluctuations on
super-Hubble scales in the contracting phase. To connect these
fluctuations to curvature fluctuations in the expanding phase, we need to
match the perturbations at the bounce by imposing the analog of the
Israel matching conditions \cite{Israel} on a space-like surface
\cite{Hwang, DM}. As discussed in detail in \cite{Peter, DV}, the result
depends very sensitively on the choice of the matching surface. Choosing
the matching surface to be the constant time surface yields the result
that the spectrum of the canonical variable $v$ is unchanged across the
bounce \cite{Lyth}. However, any other matching surface will allow the
contracting phase dominant mode of the metric fluctuation variable
$\Phi$ [see \eqref{pertmetric}] to seed the expanding phase dominant mode
of $\Phi$ which then seeds the dominant mode of $v$. This means that
this method, depending so heavily on the matching surface and conditions,
can yield any desired spectrum and thus lacks predictability.

The equation of motion for the variable 
\be 
u_k \, \equiv \, \frac{a}{\phi^{\prime}} \Phi_k 
\ee 
is 
\be 
u_k^{\prime \prime} + \left[ k^2 - \frac{p}{\left(1 - p\right)^2
\eta^2} \right] u_k \, = \, 0 \, . 
\ee 
For $p \ll 1$ this equation implies that the spectrum of $u$ (and hence
of $\Phi$) in unchanged on super-Hubble scales. Vacuum initial
conditions for the canonical variable $v$ imply, via the constraint
equations, that the initial spectrum of $u$ is 
\be 
u_k(t_i) \, \sim \, k^{-3/2} 
\ee 
which leads to a scale-invariant spectrum of $u$ and $\Phi$ fluctuations
on super-Hubble scales in the contracting phase, as initially expected
in \cite{KOST}. Taking the higher-dimensional background of Ekpyrotic
cosmology into account, one can show that a mechanism to convey the
scale-invariance of the spectrum of $\Phi$ in the contracting phase to
that of $v$ in the expanding phase naturally arises \cite{BBP} (see
also \cite{Khoury2, Khoury3, Ijjas2, Levy} for other studies of how
to obtain a scale-invariant spectrum of fluctuations in Ekpyrotic
cosmology).

A new variant of the Ekpyrotic scenario \cite{Ijjas} was recently
proposed which combines features of the original Ekpyrotic scenario with
standard inflation. This model, called the {\it anamorphic universe}, has
as a new ingredient a time dependence of masses (in this respect there
are similarities with the ``varying speed of light'' scenario of
\cite{Moffat, AAJM}). As a consequence, the model looks like standard
inflation for cosmological fluctuations, whereas matter feels an
Ekpyrotic bounce.

\subsection{String Gas Cosmology}

{\it String Gas Cosmology} \cite{BV} is based on key fundamental
principles of superstring theory, specifically on degrees of freedom and
symmetries which are absent in an effective field theory approach to
early universe cosmology. At the outset the assumption is made that all
spatial sections are finite. We will in fact assume that space is a
nine-dimensional torus.

The first key input is the Hagedorn spectrum of string states, which
leads to a maximal temperature which a gas of strings in thermal
equilibrium can achieve, the ``Hagedorn temperature'' $T_\mathrm{H}$
\cite{Hagedorn}. Closed strings in fact contain three types of modes:
momentum modes which label the center of mass motion of strings, winding
modes which label the number of times which the string winds space.
Finally there are the oscillatory modes of the strings which have a
Hagedorn spectrum. In an effective field theory description, one only
keeps the momentum modes.

\begin{figure}[t] \begin{center}
\includegraphics[scale=0.15]{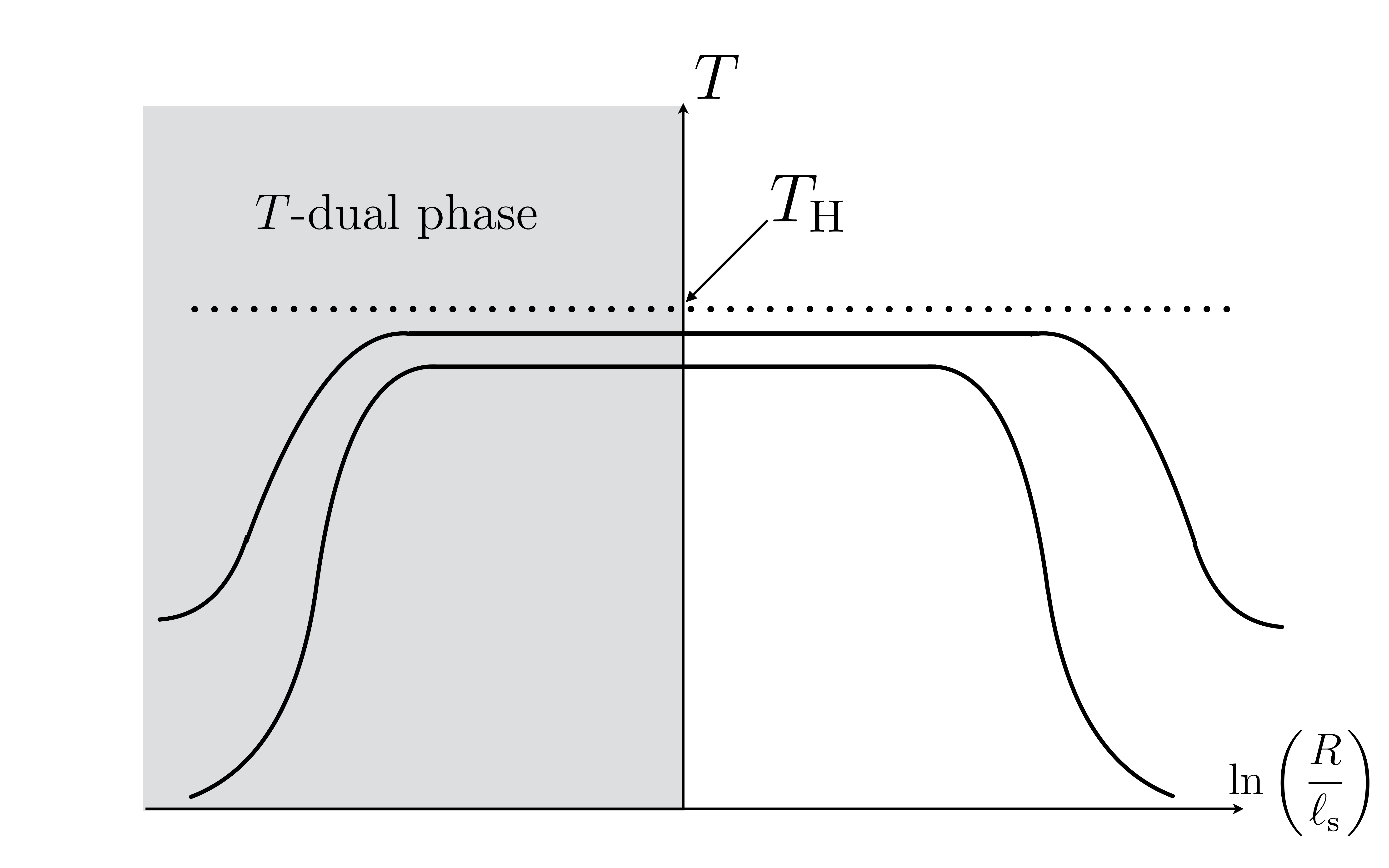}
\caption{Temperature $T$ of a gas of closed strings in a box of radius
$R$ as a function of radius. The temperature never exceeds the Hagedorn
temperature $T_\mathrm{H}$. The extent of the plateau of the $T(R)$
curve depends on the total entropy of the system - the larger the
entropy the wider the plateau.}
\label{jirofig1}
\end{center} \end{figure}

Let us now consider a box of strings in thermal equilibrium and
initially with large radius. Since the energy of the momentum modes is
quantized in units of $1/R$, where $R$ is the radius of a toroidal
section, the momentum modes are light and most of the energy of the
thermal bath will be in these modes. As we decrease the radius $R$, the
momentum modes become heavier. In thermal equilibrium, the energy will
gradually flow into the oscillatory modes (whose energy is independent
of $R$) and into the winding modes whose energy is quantized in units of
$R$ and hence become light as $R$ decreases. The temperature initially
increases as $R$ decreases, but then levels off as the Hagedorn
temperature $T_H$ is approached. Once $R$ becomes very small, the energy
all flows into the winding modes and the temperature $T(R)$ decreases
again (see Fig.~\ref{jirofig1}).

Here we in fact encounter the second key aspect of string theory, namely
the T-duality symmetry of the spectrum of string states which states
that the spectrum of string states is unchanged under the transformation
\be 
R \, \rightarrow \, \frac{\ell_\mathrm{s}^2}{R} \, , 
\ee 
where $\ell_\mathrm{s}$ is the string length scale. This symmetry is
obeyed by the perturbative string interactions and is generally taken to
be a symmetry of non-perturbative string theory (see \cite{Pol} for a
textbook review).

It is now obvious that if ${\rm ln}(R /\ell_\mathrm{s})$ increases from
$- \infty$ to $+ \infty$, the temperature-time curve will correspond to a
temperature bounce. As argued in \cite{BV}, a physical way to measure
distance for $R \gg \ell_\mathrm{s}$ is in terms of the position
operator dual to the light momentum modes, and for $R \ll
\ell_\mathrm{s}$ it is in terms of the position operator dual to the
light winding modes. In terms of this physical position operator, we
have a bouncing cosmology as the mathematical variable ${\rm ln}(R /
\ell_\mathrm{s})$ increases from $- \infty$ to $+ \infty$. Thus, string
gas cosmology can be viewed as a bouncing cosmology.

Let us consider democratic initial conditions with all nine space
dimensions small and wound by the winding modes of the equilibrium
string gas at a temperature close to the Hagedorn temperature. The
winding modes initially prevent any of the spatial dimensions to grow.
In order for space to expand, the winding modes must be able to
annihilate. This requires the intersection of string world sheets. Since
string worlds sheets - in the absence of long range forces - have
vanishing probability to intersect in more than three spatial
dimensions, only three dimensions are able to ``effectively
decompactify'' and become large \cite{BV}. This argument has been been
confirmed by numerical simulations \cite{Mairi}, although there are
important caveats \cite{Brian, Andrew}. Thus, it appears completely
natural from the point of view of string theory that there are only
three large dimensions of space even in a theory which is defined in
nine space dimensions.

The fact that string gas cosmology takes the role of winding modes into
account allows a simple stabilization of size \cite{size} and shape
\cite{Edna} moduli. Introducing gaugino condensation also allows the
stabilization of the dilaton \cite{Andrew2} without destabilizing the
size and shape moduli. In addition, gaugino condensation triggers high
scale supersymmetry breaking \cite{Wei}.

\begin{figure}[t]
\begin{center}
\includegraphics[scale=0.3]{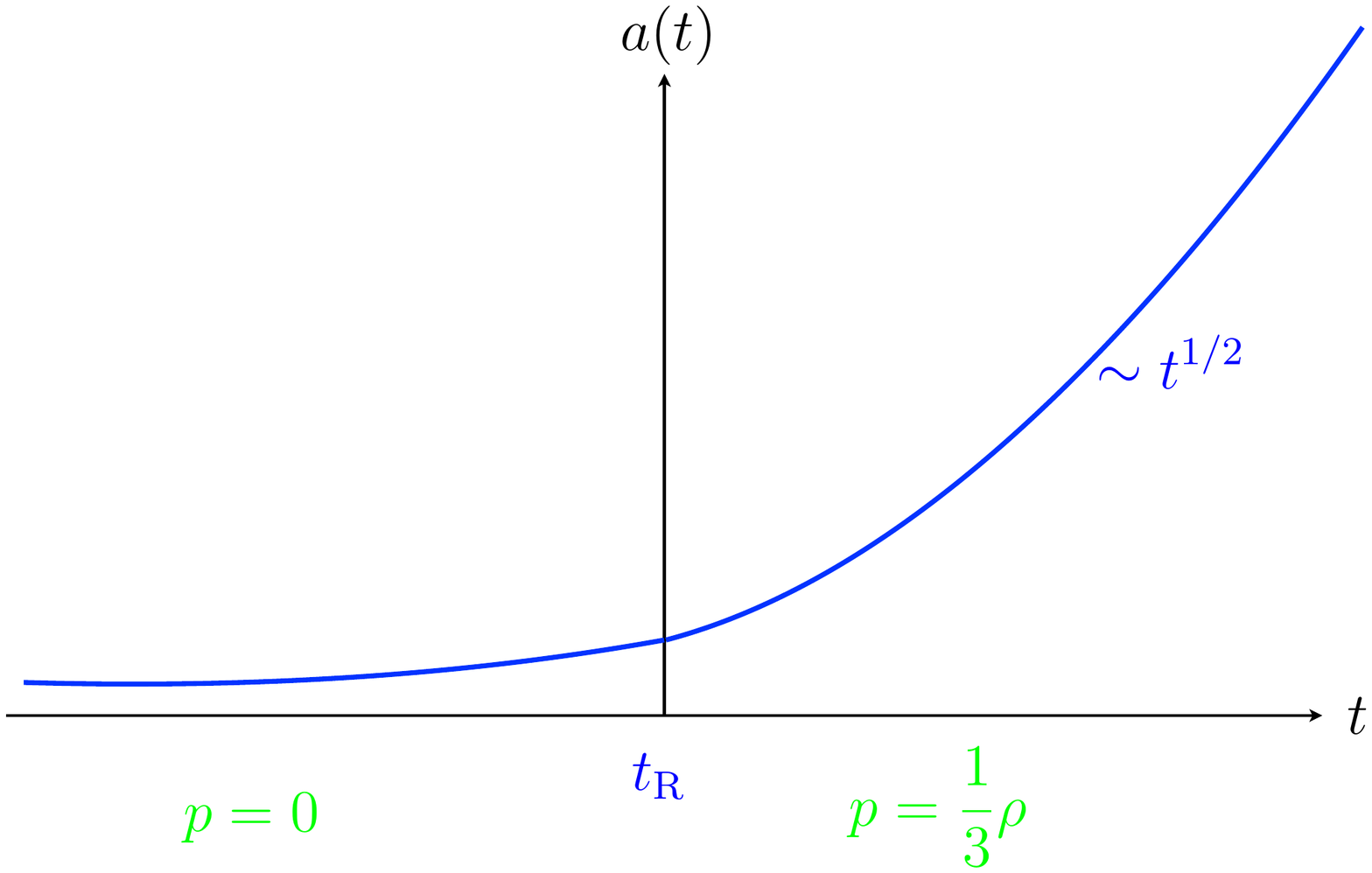}
\caption{Sketch of the evolution of the scale factor (vertical axis) as a
function of time (horizontal axis) in string gas cosmology. The time
$t_\mathrm{R}$ is the end of the Hagedorn phase and the beginning of the
radiation phase of expansion.}
\label{timeevol}
\end{center}
\end{figure}

If the universe contains a large amount of entropy, then the range of
values of $R$ for which the temperature is close to the Hagedorn value
is wide (see Fig.~\ref{jirofig1}). Hence, it is natural to suppose that
the phase with temperature close to the Hagedorn temperature will be a
long one, sufficiently long to maintain thermal equilibrium over a
distance scale which is larger than the physical length of the comoving
scale which corresponds to our current Hubble radius. The time evolution
of the effective cosmological scale factor is sketched in
Fig.~\ref{timeevol}. In this figure, the horizontal axis is time, with
$t = t_\mathrm{R}$ being the time of the phase transition when many of
the string winding modes annihilate and the equation of state becomes
that of a radiation fluid. The vertical axis represents the scale
factor.

\begin{figure}[t] \begin{center}
\includegraphics[scale=0.3]{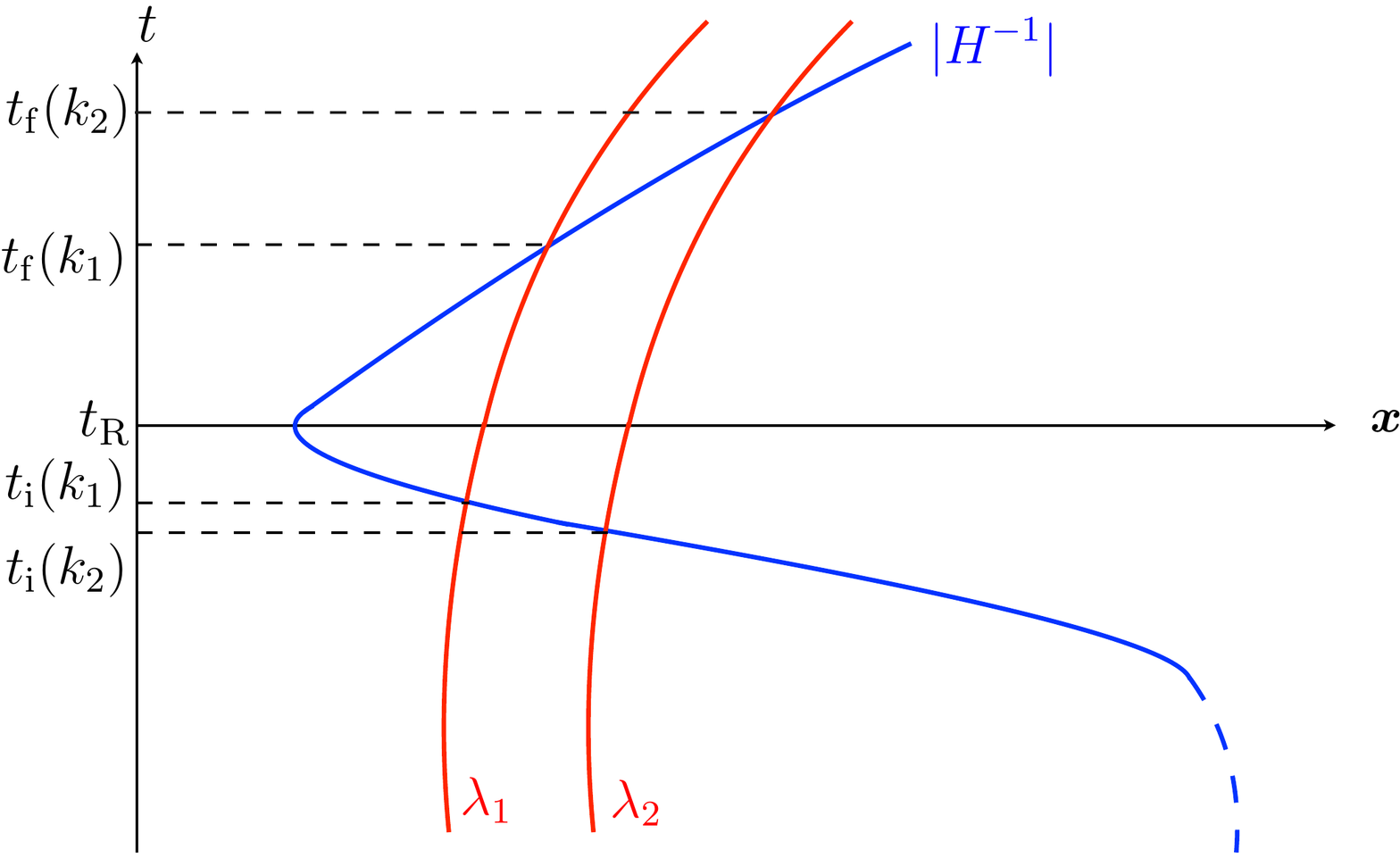}
\caption{Space-time sketch of string gas cosmology. The vertical axis is
time, with the time $t_\mathrm{R}$ being the end of the Hagedorn phase.
The horizontal axis represents physical distance. Shown are the Hubble
radius $H^{-1}(t)$ and the wavelengths of two typical modes (the curves
labelled by $\lambda_1= a/k_1$ and $\lambda_2=a/k_2$). For each of these
modes, the times when the modes exit and re-enter the Hubble radius are
indicated.} \label{spacetimenew} \end{center} \end{figure}

The corresponding space-time sketch is shown in Fig.~\ref{spacetimenew}.
Now the vertical axis is time, and the horizontal axis indicates
physical length. The initial phase $t < t_\mathrm{R}$ is quasi-static
and hence the Hubble radius goes to infinity. The physical wavelength of
fluctuation modes is constant in this early Hagedorn phase, and smaller
than the Hubble radius. Hence, a causal generation mechanism for
fluctuations is possible. If the Hagedorn temperature is of the order of
the scale of particle physics Grand Unification, then the physical
wavelength of fluctuations which are observed today is similar to the
length these modes would have at the bounce point in a matter bounce
whose energy scale is that of Grand Unification, and also similar to the
length they would have at the end of a period of inflation. Hence, the
wavelengths are many orders of magnitude larger than the Planck length,
and hence far removed from the trans-Planckian zone of ignorance.

The string gas cosmology mechanism of structure formation developed in
\cite{NBV} (see the reviews in \cite{SGCrevs}) is based on the
assumption that our four-dimensional space-time begins in a
quasi-static phase of thermal equilibrium. In this setup, thermal
fluctuations will be much more important than vacuum perturbations.
Hence, it is natural to assume that fluctuations originate as thermal
fluctuations of a gas of strings.

The idea of the computation of cosmological fluctuations \cite{NBV} and
gravitational waves \cite{BNPV} in string gas cosmology is to first
compute matter correlation functions on sub-Hubble scales in the
Hagedorn phase, and to use the Einstein constraint equations to induce
the metric perturbations scale by scale at the time when the scale exits
the Hubble radius [time $t_\mathrm{i}(k)$ in Fig.~\ref{spacetimenew}].
In particular, the cosmological fluctuations are given by 
\be \label{scalarexp} 
\langle|\Phi(k)|^2\rangle \, = \, 16 \pi^2 \GN^2
k^{-4} \langle\delta T^0{}_0(k) \delta T^0{}_0(k)\rangle \, , 
\ee 
$\GN$ being Newton's constant, in terms of the energy density
perturbations, and the gravitational waves are given by 
\be 
\label{tensorexp} \langle|h(k)|^2\rangle \, = \, 16 \pi^2 \GN^2
k^{-4} \langle\delta T^i{}_j(k) \delta T^i{}_j(k)\rangle \, 
\ee 
(with $i \neq j$) in terms of the off-diagonal pressure perturbations. 
In both cases, the brackets indicate thermal expectation values.

For a thermal gas, the energy density fluctuations on a length scale $R$
are given by the specific heat capacity $C_V$, where $V$ is the volume
associated with $R$: 
\be \label{cor1b} 
\langle \delta\rho^2 \rangle \,  = \,  \frac{T^2}{R^6} C_V \, . 
\ee 
From the partition function of closed string thermodynamics (see e.g.
\cite{Jain}) it can be shown that the specific heat capacity is (see
\cite{NBV, Nayeri, SGCrevs} for more details) 
\be \label{specheat2b} 
C_V  \, \approx \, 2
\frac{R^2/\ell_\mathrm{s}^3}{T \left(1 - T/T_\mathrm{H}\right)}\, , 
\ee
where $T$ is the temperature of the gas (which in the Hagedorn phase is
slightly lower than the Hagedorn value $T_\mathrm{H}$). Hence, the power
spectrum $\mathcal{P}_\Phi(k)$ of scalar metric fluctuations becomes 
\begin{equation}
\begin{split}
\mathcal{P}_\Phi (k)  \equiv &  \frac{1}{2 \pi^2} k^3
|\Phi(k)|^2 \\ =& \, 8 \GN^2 k^{-1} \langle|\delta \rho(k)|^2\rangle 
 \\ =& \, 8 \GN^2 k^2 \langle(\delta M)^2\rangle_R 
\\ =& \, 8 \GN^2 k^{-4} \langle(\delta \rho)^2\rangle_R  \\
=& \, 8 \GN^2 \frac{T}{\ell_\mathrm{s}^3} \left( 1 -
\frac{T}{T_\mathrm{H}} \right)^{-1} \, ,  
\end{split}
\label{power2} 
\end{equation}
where in the first step we have used (\ref{scalarexp}) to replace the
expectation value of $|\Phi(k)|^2$ in terms of the correlation function
of the energy density, and in the second step we have made the
transition to position space. In the above, the temperature $T$ is the
temperature at the time when the scale being considered exits the Hubble
radius. Since the temperature is approximately constant in the Hagedorn
phase, the spectrum of cosmological perturbations is approximately
scale-invariant. Looking a bit more closely, we see that smaller length
scales exit the Hubble radius slightly later, when the temperature is
slightly lower. Hence, the string gas cosmology structure formation
scenario automatically generates a small red tilt in the spectrum, like
simple inflation models.

As realized in \cite{BNPV}, the spectrum of gravitational waves is also
approximately scale-invariant. However, a small {\it blue} tilt is
predicted, whereas all inflation models based on General Relativity and
matter obeying the NEC predict a small red tilt. The gravitational wave
spectrum is given by \cite{BNPV} 
\be \label{tresult} 
\mathcal{P}_h (k) \, \sim \,
\left(\frac{\ell_\mathrm{Pl}}{\ell_\mathrm{s}}\right)^4 \left(1 -
\frac{T}{T_\mathrm{H}}\right)\ln^2 \left[\frac{1}{\ell_\mathrm{s}^2 k^2}
\left(1 - \frac{T}{T_\mathrm{H}}\right)\right] \, . 
\ee 
The predicted blue tilt of the tensor modes comes from the factor $(1 -
T/T_\mathrm{H})$ in (\ref{tresult}), in the same way that the red tilt
of the scalar modes comes from the factor $(1 - T/T_\mathrm{H})^{-1}$ in
(\ref{power2}).

Note that if the string scale is taken to be the preferred one from
heterotic string phenomenology (see \cite{GSW} for a textbook
treatment), then the amplitude of the resulting cosmological
perturbations has the right order of magnitude to match observations.

The three key inputs required in order to obtain a scale-invariant
spectrum of cosmological perturbations from an early thermal phase are
1) a holographic scaling of the heat capacity, 2) a quasi-static early
phase ending with a phase transition to the usual radiation phase of
expansion, and 3) the applicability of the perturbed Einstein equations
for infrared fluctuation modes.

\section{Realizations of a Bouncing Phase}

\subsection{Introductory Remarks}

According to the Hawking-Pensose theorems \cite{Hawking}, an initial
cosmological singularity inevitably arises in a homogeneous and
isotropic model if we work in the context of General Relativity and if
the matter which couples minimally to gravity obeys the Null Energy
Condition (NEC). In order to obtain a bouncing cosmology there are thus
several routes. One can work in the context of Einstein gravity but
introduce matter which violates the NEC \cite{Peter:2002cn}.
There are several dangers in
taking this route. First, one must avoid instabilites such as ghost
\cite{ghost} and gradient instabilities. Secondly, one faces the
challenge of embedding such a model in the framework of an ultraviolet
complete theory of matter and gravity \cite{Adams}.

On the other hand, any quantum theory of gravity leads to terms in the
effective gravitational action which go beyond the Einstein-Hilbert
term. Hence, it may be more promising to attempt to realize bouncing
cosmologies in the context of modified theories of gravity. This is the
topic of the third subsection.

Ultimately, however, it would best best if a bouncing scenario could be
the result of an ultraviolet complete theory of all four forces of
nature, such as superstring theory. Attempts in this direction will be
briefly reviewed in the final subsection.

Another way to classify bouncing models is whether they are singular
from the point of view of an effective field theory (such as the
Pre-Big-Bang or initial Ekpyrotic scenarios) or non-singular. We will
give examples of both.

\subsection{Bouncing Cosmologies from Modified Matter}

The simplest way to obtain a nonsingular bouncing model with modified
matter is to introduce a field with opposite sign kinetic energy term,
i.e. a ``ghost'' field, and to arrange that during the contraction phase
the absolute value of the ghost field energy density grows relative to
that of regular matter. This is the {\it quintom cosmology} scenario
\cite{quintom}. For example, the regular matter can be described by a
perfect fluid \cite{Peter:2002cn} or by a massive scalar $\phi$ field
whose time-averaged energy density scales as $a^{-3}$, and ghost matter
by a free scalar field $\psi$ with opposite sign kinetic term whose
energy density is dominated by the ${\dot \psi}^2$ term and whose energy
density hence scales as $a^{-6}$ \cite{Yifu1}\footnote{It should be
noted that in \cite{Parker:1973qd}, a regular scalar field was used in
conjunction with positive spatial curvature to avoid the primordial
singularity through a bounce: a special state was assumed, with very large
occupation number, that was leading to a direct violation of the NEC.}.
The action is 
\be 
S = \int \left[ \frac{1}{2} \partial^{\mu} \phi \partial_{\mu}
\phi -V(\phi) - \frac{1}{2} \partial^{\mu} \psi \partial_{\mu} \psi
\right] \sqrt{-g} \, \dd ^4 x\,  , \label{AW}
\ee 
with potential 
\be 
V(\phi) \, = \, \frac{1}{2} m^2 \phi^2 \, . \label{mphi}
\ee 
The non-singular bounce takes place when the energy densities of $\phi$
and $\psi$ become equal.

A specific realization of the quintom cosmology scenario can be
obtained in the {\it Lee-Wick model} \cite{LW}, a model in which the
quadratic divergences in loop amplitudes are canceled by adding
a ghost field. Bouncing cosmologies can easily be obtained making
use of the Lee-Wick construction \cite{LWbounce}. However, the
ghost problem does not appear to be solved. The ghost problem is
absent, however, in the {\it conformal cosmology} of \cite{BSTu}
(see also \cite{Gielen:2015uaa}),
a model in which the scalar sector consists of two fields with an
indefinite metric on the space of kinetic terms, but in which the ghost
is absent because the ghost degree of freedom can be eliminated
by a gauge choice. At the homogeneous and isotropic level, the
models of \cite{BSTu} have bouncing cosmological solutions.

At the classical level, the background equations of motion can be
studied and show the existence of a smooth bounce. Cosmological
fluctuations can also be evolved explicitly through the bounce phase,
with the result that on large scales (wavelength larger than the
duration of the bounce phase) the spectrum of fluctuations before and
after the bounce has the same spectral index. In particular, a
scale-invariant spectrum of cosmological flluctuations set up during the
matter phase of contraction retains its scale-invariant shape after the
bounce \cite{Yifu1}.

On the other hand, at the quantum level quintom models suffer from a
ghost instability \cite{ghost} - the vacuum can decay into pairs of
regular and ghost particles.

An improved nonsingular bounce can be obtained by the ``ghost
condensate'' mechanism \cite{ghostcond} (see also \cite{Kbounce}). 
Ghost condensation is the
kinetic sector analog of the Higgs mechanism in the potential sector. In
the case of the Higgs field $\phi$, the theory has a tachyon when
expanded about $\phi = 0$, but not when expanded around the true minimum
of the potential. In the ghost condensate scenario, we introduce a
non-trivial kinetic term $K(X)$ in the action 
\be 
S \, = \, \int  \left[ K(X) - V(\phi) \right] \sqrt{-g}\, \dd^4 x \, , 
\ee 
where $X$ is the kinetic term 
\be 
X \, = \, \partial^{\mu} \phi \partial_{\mu} \phi \, , 
\ee 
and we choose the kinetic function $K(X)$ such that the Lagrangian is
ghost-like when expanded about $X = 0$, but has the regular sign when
expanded about the true minimum of $K(X)$. Thus, in the ghost condensate
phase there is no perturbative ghost instability.

Applied to bouncing cosmology, we can take \cite{Chunshan} 
\be 
K(X) \, = \, \frac{1}{8} \left( X - c^2\right)^2 \, , 
\ee 
where $c$ is some positive constant. We consider a potential $V(\phi)$
which decays exponentially as $\phi \rightarrow \pm \infty$ and is
positive at $\phi = 0$. As initial conditions we consider a contracting
universe with $\phi = ct$ as $t \rightarrow - \infty$, i.e. with the
field in the ghost condensate state. As $t$ approaches zero, the
potential starts becoming important and slows down the field, thus
leading the ghost field to obtain negative kinetic energy. The bounce
point is reached when the ghost kinetic energy cancels the potential
energy.

At a classical level, the dynamics is well behaved. One can show that,
like in the case of the quintom bounce, on large scale a scale-invariant
spectrum of cosmological fluctuations passes through the bounce without
a change in the spectral index. At a perturbative quantum level the
model is also safe, but there is the worry \cite{Adams} that it is not
possible to embed a ghost condensate Lagrangian into an ultraviolet
complete theory of physics.

An additional problem is a gradient instability from which the model
suffers. This problem can be cured by replacing the ghost condensate
Lagrangian by a Galileon Lagrangian \cite{Galileon} and considering
Galileon bounces \cite{GalBounce}.

There are also other avenues of obtaining a bouncing cosmology from
modified matter which have been explored. For example, the possibility of
obtaining a bounce from a Fermion condensate in the presence of a
non-trivial coupling of General Relativity to fermionic fields has been
explored in \cite{Stephon}. Another idea is to use coupled scalar
tachyons \cite{Edna3}. In a very different spirit, it has been realized
that the instability of the Standard Model Higgs field can be used
\cite{Youping} to generate a phase of Ekpyrotic contraction (but one
needs an additional ingredient to obtain the actual bounce).

\subsection{Bouncing Cosmologies from Modified Gravity}

In any approach to quantum gravity, terms in the gravitational action
which are additional to the usual Einstein-Hilbert term are predicted.
These terms may be higher derivative in the effective
gravitational action \cite{higher}, and they may be non-local
\cite{nonlocal}. Assuming that the matter theory is a standard one given
in terms of an energy-momentum tensor $T_{\mu \nu}$, then the effective
equations of motion including leading quantum correction may be
written as 
\be \label{EoM5} 
{\cal D}_{\mu \nu}(g_{\alpha \beta}) \, = \, 8 \pi \GN
T_{\mu \nu} \, , 
\ee 
where ${\cal D}_{\mu \nu}$ is an operator which contains the Einstein
tensor $G_{\alpha \beta}$ as leading term, but in addition the quantum
gravitational higher derivative and/or non-local terms. We can extract
the leading term and write 
\be 
{\cal D}_{\mu \nu}(g_{\alpha \beta}) \, = \, G_{\mu \nu} +
{\tilde{\cal D}}_{\mu \nu}(g_{\alpha \beta}) \, . 
\ee 
Equivalently, we can write (\ref{EoM5}) as 
\be 
G_{\mu \nu} \, = \, 8 \pi \GN T^{\rm eff}_{\mu \nu} \, 
\ee 
with 
\be 
T^{\rm eff}_{\mu \nu} \, = \, T_{\mu \nu} - \frac{1}{8 \pi \GN} 
{\tilde{\cal D}}_{\mu \nu}(g_{\alpha \beta}) \, . 
\ee 
It is now easy to see that even with a matter energy-momentum tensor
which obeys the NEC, it is possible to obtain a bounce if the
gravitational contribution ${\tilde{\cal D}}_{\mu \nu}(g_{\alpha
\beta})$ is such that the effective energy-momentum tensor violates the
NEC.

One example where the idea can be realized is Horava-Lifshitz gravity
\cite{HL}, an approach to quantum gravity which gives up  general
space-time diffeomorphism invariance and maintains only spatial
diffeomorphism invariance plus a new space-time anisotropic scaling
symmetry 
\be 
t \, \rightarrow \, \ell^3 t \,\, \ \ \ {\rm and} \,\, \ \ \ x^i \,
 \rightarrow \, \ell x^i \, , 
\ee 
where $\ell \in \mathbb{R}$ is a scaling parameter. One can determine
the most general gravitational action which is consistent with the above
symmetries and which is power-counting renormalizable. The resulting
gravitational field equations contain a new gravitational contribution
${\tilde{\cal D}}_{\mu \nu}(g_{\alpha \beta})$ which, when taken to the
right hand side of the gravitational field equations, has the form of
dark radiation with negative effective energy density whose energy
density scales as $a^{-4}$ in a contracting FLRW
universe. The coefficient of the dark radiation depends on spatial
curvature. We only obtain dark radiation for positively or negatively
curved spaces, but not in the spatially flat case. Provided that regular
matter scales in time less fast than $a^{-4}$, e.g. if it has a
matter-dominated equation of state, then eventually, the contribution of
dark radiation will become equal in magnitude to that of regular matter,
and a cosmological bounce will occur \cite{HLBounce} (see also
\cite{HLcosmo}).

The main worry about extensions of General Relativity is the issue of
stability of metrics we know should be stable such as the FLRW
or the Schwarzschild metrics. For cosmological backgrounds, we
have to worry about ghost-like or tachyonic instabilities of the
linearized perturbations. This issue was first studied in
\cite{HLpert1}, and then in the context of a bouncing cosmology in
\cite{HLpert2}. There are several versions of Horava-Lifshitz gravity.
One of them is called the ``projectable version'', another is the
original version \cite{HL} extended by terms which were only implicit in
the original version - the ``healthy extension'' \cite{healthy}. It was
found that in the healthy extension of Horava-Lifshitz gravity, there
are no disastrous instabilities at linear order \cite{Cer1}, but in the
projectable version there are \cite{Cer2}. In the healthy version, it
can moreover be verified that the spectral index of the cosmological
fluctuations is unchanged across the bounce \cite{HLpert2}.

Bouncing cosmologies have also been studied in the context of
$F(R)$ theories of gravity (see e.g. \cite{FRG}), in models in which the
gravitational action is supplemented by a Gauss-Bonnet term
(see e.g.  \cite{GRB}) and in $F(T)$ theories of gravity 
(see e.g. \cite{FTB}). Nonsingular cosmological solutions also
emerge from the Einstein-Cartan-Sciama-Kibble gravitational
action \cite{Desai:2015haa}. Bouncing cosmologies can also be obtained
in models in which our space-time is a brane moving through a
higher-dimensional space-time which is curved (e.g. of black hole
type) \cite{BraneBounce}, and in bouncing braneworlds \cite{Shtanov:2002mb}. 
Bouncing cosmologies are also an inevitable
consequence of the ``limited curvature gravitational action" of \cite{BMS}.

Obviously bouncing cosmologies can also be obtained if we modify
both the matter and the gravitational actions. For example, bouncing
cosmologies from negative potentials in scalar-tensor theories have
recently been studied in \cite{Boisseau:2015hqa}.

\subsection{Bouncing Cosmologies from String Theory}

String theory is a candidate ultraviolet complete theory which unifies
all four forces of nature at a quantum level. As already seen in the
subsection on string gas cosmology, it should be expected that string
theory will lead to a very different evolution of the early universe
than what is obtained in the effective point particle theory limit. At
the present time, however, string gas cosmology is based on ideas from
perturbative string theory coupled with thermodynamic arguments. A
rigorous non-perturbative treatment is lacking. Hence, it is useful to
explore other approaches to string cosmology which are more closely
related to non-perturbative physics, and which result in bouncing
cosmologies. String gas cosmology is one example, but in this subsection
we will focus on a couple of attempts which can be formulated in terms
of an effective background field theory.

One such approach is the {\it S-brane bounce} \cite{Sbrane}, which
is based on a certain class of Type II superstring backgrounds
\cite{backgrounds} for which the string partition function has a
temperature duality 
\be 
T \, \rightarrow \, \frac{T_\mathrm{c}^2}{T} \, 
\ee 
where $T_\mathrm{c}$ is a critical temperature whose value is related to
the string scale.

We begin the evolution in a contracting universe starting with a very
large value of the temperature $T \gg T_\mathrm{c}$ (which physically
corresponds to a very small value of the effective temperature). As the
universe contracts, $T$ decreases (which means that the effective
temperature increases). Once the effective temperature exceeds the
supersymmetry breaking scale, we can describe the background cosmology
using dilaton gravity. When the temperature reaches the value $T =
T_\mathrm{c}$, a tower of string states which for $T \neq T_\mathrm{c}$
are massive (of the string scale) become massless and have to be
included in the low energy effective action. Hence, the effective action
for the low mass modes of matter contains a term localized on the
spatial hypersurface $T = T_\mathrm{c}$. This object is the space-like
analog of a D-brane in string theory, or of the zero width limit of a
topological defect in field theory. Hence, we call this term an {\it
S-brane} (see also \cite{Sbrane2} for another approach to S-branes in
string theory).

The equation of state of an S-brane has vanishing energy density $\rho =
0$ and negative pressure $p < 0$ and hence yields a term which violates
the NEC, thus allowing for a bouncing solution. As $T$ approaches the
value $T = T_\mathrm{c}$ from above, the universe is contracting, but
after the S-brane is encountered as $T = T_\mathrm{c}$ the universe
starts to re-expand, with decreasing $T$, as shown explictly in
\cite{Sbrane}. The evolution of cosmological perturbations in this
background was studied in \cite{Sbrane3} where it was shown that a
scale-invariant spectrum of curvature fluctuations passes through the
bounce unchanged.

The second approach which we will mention here is based on the AdS-CFT
correspondence \cite{Malda}, a conjecture which relates string theory on
an AdS (anti-de-Sitter) space-time to a conformal field theory (CFT)
living on its boundary. This proposal can give a non-perturbative
definition of string theory on AdS since we have a non-perturbative
description of the boundary CFT.

To apply the AdS/CFT correspondence to a bouncing cosmology, we consider
a time-dependent deformation of AdS which asymptotically for $t
\rightarrow \pm \infty$ looks like pure AdS but has growing bulk
curvature as $t \rightarrow \pm 0$. Mapped onto the CFT living on the
boundary, we obtain a time-dependent gauge coupling constant which tends
to zero as $t \rightarrow \pm 0$. This leads to the hope that the
boundary field theory time evolution can be smoothly continued through
$t = 0$, and that this continuation can then be used to construct a bulk
space-time in the future of $t = 0$ \cite{CHT} (see also \cite{HH} for
earlier work). We would thus be able to pass through the bulk
gravitational singularity (the singular cosmological bounce point) in a
unique way by mapping the physics onto a well-defined boundary theory.

Whereas this procedure works at the level of the background cosmology,
singularities on the gauge theory side arise when cosmological
fluctuations are introduced. This has recently been studied \cite{us} in
the case of a deformation of AdS obtained by a time-dependent dilaton
\cite{Sumit} (see also \cite{Pei-Ming}). The singularity in the boundary
theory takes the form of a branch cut in one of the fluctuation modes,
and in divergent total particle production \cite{us}. Hence, we need to
impose time cutoffs at $t = \pm \xi$, where $\xi$ is set, e.g., by the
bulk curvature obtaining the Planck value. In this case, the cosmological
perturbations can be evolved on the boundary in a more unique way than
they could in the bulk, where the ambiguity of the space-like matching
surface is a major problem.

Concretely, the application of the AdS/CFT correspondence to
cosmological fluctuations is now the following. We begin with bulk
fluctuations in the contracting branch of the deformed AdS. When the
bulk curvature becomes strongly coupled at the time denoted by $-
t_\mathrm{c}$, we project the fliuctuations onto the boundary gauge
theory. We then evolve the boundary fluctuations through $t = 0$ until
the time $+ t_\mathrm{c}$ when the bulk becomes weakly coupled again. At
that time we reconstruct the bulk fluctuations.

The analysis of \cite{us} shows that on large scales, the spectrum of
bulk fluctuations has the same slope at $t = - t_\mathrm{c}$ and $t =
t_\mathrm{c}$. The amplitude, however, grows by a factor which diverges
as $\xi$ approaches zero.

String gas cosmology is a scenario based on fundamental principles
of superstring theory which can yield a bouncing cosmology, as
discussed earlier. The Ekpyrotic scenario is, as well, originally
motived by ideas from superstring theory. Orbifold models
have also been studied as a means for obtaining nonsingular
cosmological solutions in \cite{Cornalba:2002fi}. For another recent
construction of a bouncing cosmology from string theory see \cite{Ednanew}.

It is also possible that bouncing cosmologies can arise from alternative
approaches to quantum gravity. Loop quantum gravity is the prime
example. In fact, in loop quantum cosmology it can be shown that
the cosmological singularity can be avoided at the quantum level,
and that bouncing cosmologies are possible \cite{LQbounce}.
For a specific construction of a ``matter bounce" in the context of
loop quantum cosmology see \cite{Edward2}.
Bouncing cosmological solutions also emerges from the
{\it Group Field Theory} approach to gravity \cite{deCesare:2016axk}.

\section{Observational Signatures}

Alternative models of early universe cosmology must be able to explain
all of the existing data and must be consistent with the current
constraints. It is also very important that they make new predictions
with which they can be differentiated from the current and widely
accepted inflationary paradigm. Given the wealth of data which is
expected in the near future from new telescopes this is an ideal time to
consider such predictions. They could concern new observational windows
such as CMB $B$-mode polarization maps, 21cm redshift surveys and
gravitational wave astronomy, or vastly improved data from radio and
optical telescopes. Below we mention a couple of examples of how new
observational windows will allow us to differentiate the predictions of
various bouncing cosmological models with those of the inflationary
scenario. We will discuss higher order effects such as non-Gaussianities,
the tensor spectrum and its relations with the observed scalar component,
and the running of the scalar spectrum (see also\cite{Chowdhury:2015cma}
for a discussion of tensor non-Gaussianities in the matter bounce).

\subsection{Non-Gaussianities}

Bouncing models contain many interaction terms ensuring that the bounce 
(and the previous phases) take place. These, in turn, lead to an
extra contribution to the 3-point function which is given by the commutator
\begin{equation}
\langle \zeta(\bm{k}_1) \zeta(\bm{k}_2) \zeta(\bm{k}_3) \rangle = i \int _{t_\mathrm{G}}^t
\langle \left[ \zeta(\bm{k}_1) \zeta(\bm{k}_2) \zeta(\bm{k}_3), \mathcal{L}_\mathrm{int} (t')
\right] \rangle \dd t', \label{3pt}
\end{equation}
where $\mathcal{L}_\mathrm{int}$ contains the relevant interaction
terms, and the curvature perturbations $\zeta(\bm{k}_i)$ are evaluated
at time $t$; $t_\mathrm{G}$ is an initial time at which no 
non-Gaussianity is yet present \footnote{In this subsection we
use the convention that the mass dimension of $\zeta(k)$ is $-3$ and
not $-3/2$ as we did earlier in this review.}. The interaction Lagrangian
$\mathcal{L}_\mathrm{int}$ needs be calculated at least to third order
in the perturbation $\zeta$, leading to a complicated function of
$\zeta$ whose details can be found in \cite{Maldacena:2002vr} where it
was first calculated.

One then usually expresses the three-point function as
\begin{equation}
\langle \zeta(\bm{k}_1) \zeta(\bm{k}_2) \zeta(\bm{k}_3) \rangle = \left(2\pi\right)^7
\delta^{(3)} \left( \bm{k} \right) 
\frac{\mathcal{P}^2_\zeta}{\displaystyle \prod_i k_i^3}\mathcal{A},
\label{3ptf}
\end{equation}
where $\bm{k}=\sum_i \bm{k_i}$ and $\mathcal{A}$ is a shape function,
depending on the wave vectors
$\bm{k}_i$. In Eq.~\eqref{3ptf}, the quantity $\mathcal{P}^2_\zeta$
is not the actual curvature perturbation spectrum but rather
represents a sum of permutations of the 3 different wavenumbers
involved, namely 
\be
\mathcal{P}^2_\zeta = \mathcal{P}_\zeta\left(
k_1\right) \mathcal{P}_\zeta\left( k_2\right) + \mathcal{P}_\zeta\left(
k_2\right) \mathcal{P}_\zeta\left( k_3\right) + \mathcal{P}_\zeta\left(
k_1\right) \mathcal{P}_\zeta\left( k_3\right) \, . 
\ee
In general, one
restricts attention to the so-called non linearity parameter
$f_\mathrm{NL}$, defined in real space by
\begin{equation}
\zeta = \zeta_\mathrm{gauss}\left( \bm{x}\right) + \frac35 f_\mathrm{NL} 
\zeta_\mathrm{gauss}^2\left( \bm{x}\right),
\label{fNL}
\end{equation}
where $\zeta_\mathrm{gauss}\left( \bm{x}\right)$ stands for the linear,
Gaussian part of the curvature perturbation. One then arrives at
\begin{equation}
\langle \zeta(\bm{k}_1) \zeta(\bm{k}_2) \zeta(\bm{k}_3) \rangle =
\frac{3(2\pi)^7}{10}\delta^3\left(\bm{k}\right)
f_\mathrm{NL} \mathcal P_\zeta^2\frac{\sum_i k_i^3}{\Pi_jk_j^3}
\label{bispec}.
\end{equation}
Common configurations include the local shape ($k_3\ll k_1,k_2$),
the equilateral shape ($k_1=k_2=k_3$), or the folded or orthogonal shapes
\cite{Babich:2004gb,Baumann:2009ds,Battefeld:2011ut}.
Current bounds set by P{\footnotesize LANCK}, including
temperature and polarization data, are \cite{Ade:2015ava},
\begin{equation}
\begin{split}
f_\mathrm{NL}^\mathrm{local}=&0.8\pm5,\\
f_\mathrm{NL}^\mathrm{equil}=&-4\pm43,\\
f_\mathrm{NL}^\mathrm{ortho}=&-26\pm21,
\end{split}
\label{currentbounds}
\end{equation}
($68\%$ CL) which are consistent with a Gaussian (all non-linearity
parameters vanish) spectrum. Such suppressed non-Gaussianities are
consistent with the prediction in canonical, single field, slow-roll
models of inflation \cite{Maldacena:2002vr,Creminelli:2003iq}.

Bouncing cosmologies present many difficulties for the calculation of
these non-Gaussianities, as in particular many models induce large
contributions. In an inflationary setup, perturbations, and therefore
non-Gaussianities too, are produced at one instant of time, at Hubble
crossing, and then subsequently propagated mostly unchanged. Many terms
in the interaction Lagrangian $\mathcal{L}_\mathrm{int}$, e.g., those
depending on time derivatives of $\zeta_k$, are utterly negligible,
leading to a final result which never grows very large, and is in fact
largely controlled by the slow-roll parameter $\epsilon$. In a bouncing
model however, although the first mechanism is rather similar, the
subsequent evolution can be very different for two reasons. First, the
initial phase is one of contraction, during which the modes may change
with time, producing a more important growth to begin with. This implies
that the resulting contraction non-Gaussianities can be much larger than
their inflationary counterpart. This stems from the fact that the
``slow-roll'' parameter, in a contracting universe, is not a small
parameter. In  the case of the matter bounce, it was found
\cite{Cai:2009fn} that $f_\mathrm{NL}^\mathrm{local} =-\frac{35}{8}$,
comparable to the current bounds, and similarly the other
$f_\mathrm{NL}$ are expected to be negative with a magnitude of a few.

The second difference between an inflation phase and a bouncing model is
the existence of the bounce itself, which has a tendency to increase any
pre-existing non-Gaussianity. For instance, in a simple model consisting of
a single scalar field with positive spatial curvature \cite{GaoKNG}, it was
realized that the closer the bounce to de Sitter, the higher the production
of non-Gaussianities, and at a level which is such as to raise doubts on the
validity of the perturbative treatment of the bounce. Of course, one may then
argue that the problem stems from the fact that the positive spatial
curvature is crucial for the bounce to merely take place, and that therefore
this is not necessarily a valid generic result. Granted, but more recently,
it was also shown \cite{Quintin:2015rta}, using a ghost-condensate galileon
model to perform the bounce in a flat spatial background, that $f_\mathrm{NL}$
should also grow during the bounce phase. Besides, it was also shown that the
increase of the level non-Gaussianity is directly related with that of the
overall amplitude of curvature perturbation, leaving the tensor mode unchanged.
As a result, the tensor to scalar ratio, discussed in the following section, can
remain small and compatible with the current constraints at the expense having
too much non-Gaussianities. This is even a possibly fatal blow for single field
matter bounce models \cite{Quintin:2015rta}.

The non-Gaussianities in string gas cosmology, on the other hand, are
predicted to be negligible on cosmological scales \cite{SGNG}. The reason
is the following: thermal fluctuations have a specific correlation length
which is microscopic and at which scale the non-Gaussianities are thermal
and or order one. However, on larger scales the non-Gaussianities are
Poisson suppressed and hence become negligible on scales relevant for
cosmological observations. There is a caveat to this argument: if the
strings in string gas cosmology are long-lived (see e.g. \cite{CMP} for a
discussion of this point), then a network of cosmic superstrings
\cite{Witten} is predicted to survive to the present time. These strings
would approach a scaling solution of the same form as that describing
cosmic strings forming in a field theory phase transition. These cosmic
strings would leave behind distinctive stringy signals in observations,
signals with specific patterns in position space (see \cite{RHBCSrev} for
a recent review of these signals).

\subsection{Running of the Scalar Spectrum}

In simple single field inflationary models, the red tilt of the scalar spectrum is
due to the fact that the Hubble expansion rate $H$ is very slowly decreasing during the
period of inflation. In fact, the rate of decrease of $H$ is accelerating during
the slow rolling phase. This leads to the fact that the slope of the spectrum is
increasing in magnitude towards smaller wavelength. This implies that the
{\it running} of the spectrum is negative, i.e. the slope of the spectrum is smaller
at larger values of the momentum.

In the matter bounce scenario, on the other hand, the tilt of the spectrum of
cosmological perturbations is caused by the fact that the contribution of the
dark energy component is decreasing as a function of time during the
contracting phase. This implies that at large values of $k$, the spectrum converges
to a scale-invariant one, and that thus the running of the scalar spectrum is
positive. This point was recently worked out in detail in \cite{Edward3}.

Thus, a measurement of the running of the scalar spectrum would allow us
to differentiate the canonical single field inflation models from the matter bounce
scenario. An analysis of the running of the scalar spectrum in string gas
cosmology is not available at this time. 

\subsection{Tensor to Scalar Ratio}

The tensor to scalar ratio $r$ is defined by
\be 
r \, \equiv \,
\frac{\mathcal{P}_\mathcal{T}}{\mathcal{P}_\mathcal{S}} \, , 
\ee 
where $\mathcal{P}_\mathcal{T}$ and $\mathcal{P}_\mathcal{S}$ are the
power spectra of the tensor and scalar modes, respectively. In the matter
bounce scenario, both the tensor and the scalar fluctuations have a
scale-invariant spectrum, and the tensor to scalar ratio before the
bounce phase is predicted to be of order one since the scalar and tensor
modes obey the same equation of motion. During the bounce, it is possible
that the scalar modes are enhanced relative to the tensor modes. This
issue has recently been studied in \cite{Quintin:2015rta} in the case of
a particular two field scenario. Nevertheless, generically a large value
of $r$ is predicted. Hence, the value of $r$ does not provide a good
window to differentiate between inflation and the matter bounce.

In string gas cosmology, the value of $r$ is given by the ratio between
the energy density and the pressure fluctuations, as discussed in the
section on string gas cosmology. In the model of \cite{BV, NBV} there is
a consistency relation which relates $r$ to $1 - n_\mathrm{s}$. Given the
observed value of $n_\mathrm{s}$, a value of $r$ below the current
observational bounds is predicted (see \cite{BNP} for a detailed
discussion). A measurement of $r$ alone is hence also not a good way to
differentiate between string gas cosmology and inflation.

\begin{figure}[t]
\begin{center}
\includegraphics[scale=0.4]{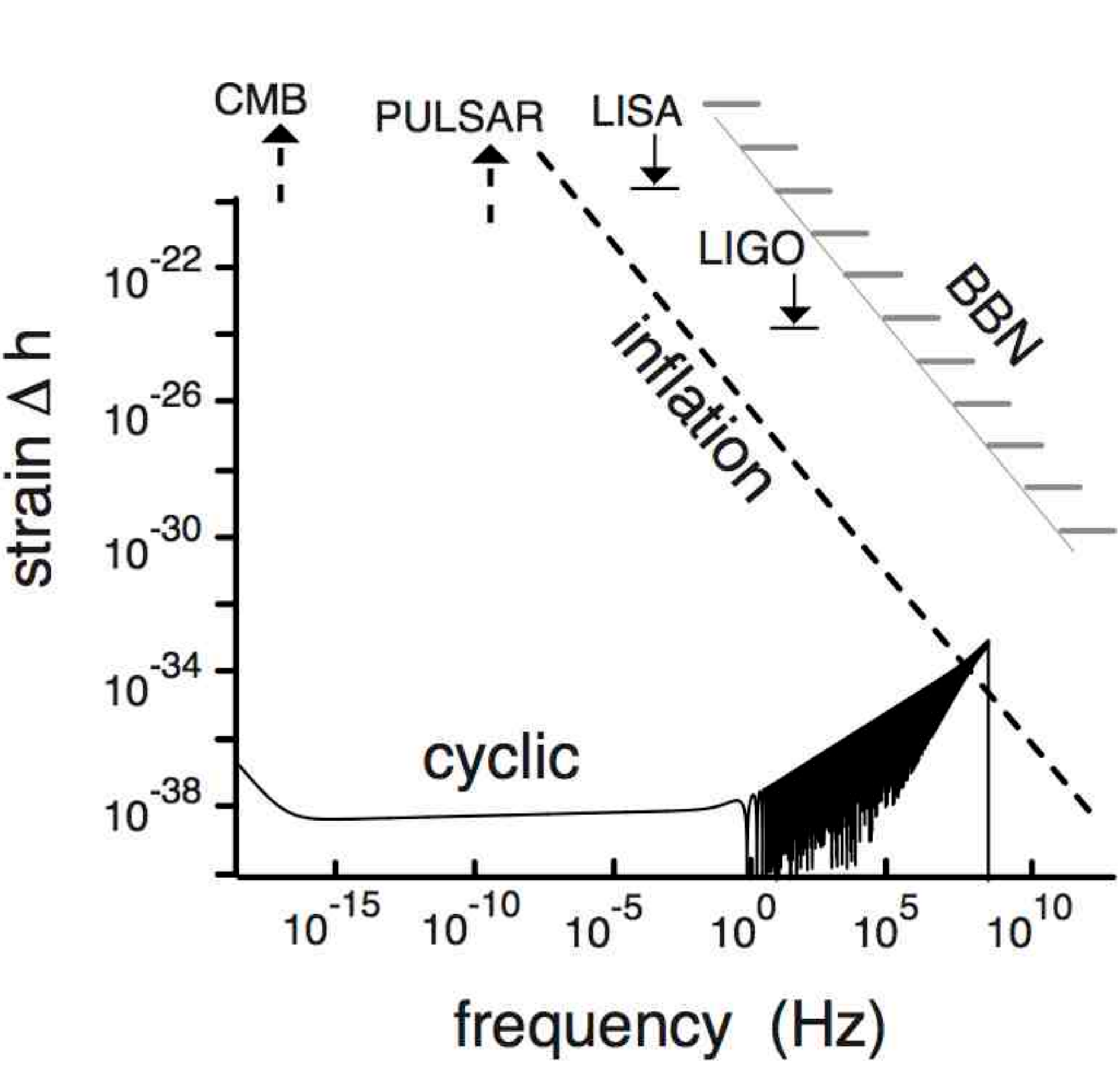}
\caption{The ekpyrotic/cyclic prediction for the power spectrum of
gravitational waves expressed in terms of the strain $\Delta h \equiv
\sqrt{\mathcal{P}_\mathcal{T}}$ as a function of the frequency. The
normalization is fixed at small wavelengths at which one assumes the
tensor contribution to be 4 orders of magnitude below the BBN
constraint. Using the blue spectrum to reconstruct this contribution on
large wavelengths, one thus finds a value of $r$ that is utterly
negligible and that is therefore predicted to be unobservable.}
\label{strain}
\end{center}
\end{figure}

The situation is completely different in the case of the 
Ekpyrotic scenario. In this case, both the adiabatic
scalar spectrum and the tensor spectrum retain their 
original vacuum slope, which means that the fluctuations
are negligible on cosmological scales. The scalar fluctuations
which are currently observed must hence be due to a 
non-standard mechanism such as entropy fluctuations,
while there is no possibility of an additional generation
mechanism for tensor modes. The quantitative analysis
begins by considering the scale factor to decrease following
Eqs.~\eqref{atp} and \eqref{aetap}, so the mode equation
\eqref{FourierEoM} gives 
\begin{equation}
\mu = \frac12\sqrt{-\pi(\eta-\eta_*)}
H^{(1)}_{\frac12-\beta}\left[-k(\eta-\eta_*)\right]
\label{muH1}
\end{equation}
where $H^{(1)}$ is the Hankel function, and $\beta\ll 1$; $\eta_*$ is
the time at which the ekpyrotic potential would reach negative infinity
if it were described at all times by the exponential shape leading to
ekpyrotic contraction. The tensor spectrum
\begin{equation}
\mathcal{P}_\mathcal{T} = \frac{k^3}{\pi^2}|h|^2
\label{PTekp}
\end{equation}
must then be transfered through the kinetic driven phase,
the bounce, and the subsequent expanding phases by means of either
matching conditions or numerical evaluation. A value for the tensor-to-scalar
ratio of $r\sim 0.1$ implies that, today, the gravitational power spectrum
should be of order $\mathcal{P}_\mathcal{T} (r\sim 0.1) \simeq 10^{-10}$
on CMB scales. The most stringent constraint on tensor modes then comes
from big-bang nucleosynthesis (BBN), during which it has to be negligible. This
implies contraints  on the amplitude at high frequencies, i.e. small wavelengths.
But the power spectrum being very blue, this also implies that on the much
longer wavelengths concerned by the CMB, the tensor contribution should
be even smaller. In fact, according to this line of thoughts, any sizeable
measurement of $r$ would immediately rule out the Ekpyrotic scenario
unless some yet-unknown mechanism is invoked. The actual calculation
was done in Ref.~\cite{Boyle:2003km}, from which we reproduce the result
in Fig.~\ref{strain}

The value of $r$ is however very highly model dependent in bouncing
cosmologies (just as it is very discriminatory in inflation
model-building \cite{Martin:2014rqa}), as illustrated by a simple
effective bounce model with two fields: considering two scalar fields
$\phi$ and $\psi$, the latter evolving in an exponential potential
reminiscent of the Ekpyrotic situation (but positive) and the former
having a negative definite kinetic energy only, i.e. \eqref{AW} with the
potential \eqref{mphi} replaced by $V=V_0 \exp\left( -\lambda
\phi/m_\mathrm{Pl}\right)$, the authors of Ref.~\cite{Allen:2004vz}
showed that during the collapse phase, one finds $r=8\lambda^2$. Since
this is the only available free parameter, it also fixes the scalar
spectral index. In order to recover a scale-invariant spectrum, one needs
to set $\lambda=\sqrt{3}$, leading to $r=24$, much in excess of the
current observational constraint.

In a matter bounce scenario with a phase of Ekpyrotic contraction used
only to wash any primordial anisotropy, the curvature fluctuations can
grow relatively large, without spoiling the perturbative nature of the
bounce, while the tensor modes remain small at all times. In such a
model therefore, the tensor-to-scalar ratio remains under control and
can be made compatible with the data \cite{GalBounce}. Similar
consideration apply to the matter bounce curvaton scenario
\cite{Cai:2011zx}.

\subsection{Tilt of the Tensor Spectrum}

All inflationary models based on Einstein gravity coupled to matter
which obeys the NEC predict a red tilt of the tensor spectrum, i.e.
$n_\mathrm{t} < 0$. The reason is that, during inflation, the magnitude of
$H$ is a decreasing function of time\footnote{Models of inflation based
on matter which violates the NEC can yield a blue spectrum, see e.g.
\cite{GalInfl}.}. The amplitude of the gravitational waves on a
particular scale $k$ is, as we have seen in Section 2.1, proportional to
the value of $H$ at the time $t_H(k)$ when the scale exits the Hubble
radius during inflation. Larger values of $k$ correspond to waves which
exit later and hence at smaller values of $H$.

In the case of simple single field slow-roll inflationary models there
is, in fact, a consistency relation which connects the values of the
tensor tilt with the tensor to scalar ratio $r$. It takes the
form \cite{LL} 
\be \label{Icons} 
n_\mathrm{t} \, = \, - \frac{r}{8} \, . 
\ee 
Whereas the sign of $n_\mathrm{t}$ is a very generic prediction of
scalar field inflation, the specific relation between $n_\mathrm{t}$ and
$r$ is not generic since by complicating the matter sector one can
change the scalar fluctuations.

In the case of the simple matter
bounce model, one obtains the consistency relation 
\be \label{Mcons} n_\mathrm{t} \, = \, (n_\mathrm{s} - 1) \, , \ee 
the reason being that both sets of fluctuations start out as vacuum
perturbations, and that the squeezing of the scalar and tensor modes is
the same. The same applies for the ekpyrotic contraction, with
$n_\mathrm{t}=3$ [see Eq.~\eqref{nt3}]. 

On the other hand, in string gas cosmology the origin of fluctuations is
thermal as opposed to quantum, and the initial amplitudes of the scalar
and tensor modes are very different. The scalar mode amplitude is set by
the value of the temperature, which slowly decreases as a function of
time, leading to a red tilt. On the other hand, the amplitude of the
tensor modes is set by the pressure anisotropies, which are proportional
to the presssure, which increases as a function of time at the end of
the Hagedorn phase, thus leading to a blue tilt of the tensor modes
\cite{BNPV}. A careful analysis shows the following string gas
consistency relation between the scalar and tensor modes \cite{BNP} 
\be \label{Scons} 
n_\mathrm{t} \, \simeq \, - (n_\mathrm{s} - 1) \, .
\ee 
Hence, measuring the value of $r$ and the tensor tilt would allow us to
discriminate between different early universe scenarios.

Tensor modes on cosmological scales leave an imprint on the spectrum of
$B$-mode CMB polarization. Scalar cosmological fluctuations cannot
produce $B$-mode polarization at linear order in perturbation theory.
Gravitational waves, on the other hand, do. The challenge for future
observations is not only to detect the amplitude of $B$-mode
polarization, but to determine the tilt. A determination of only the
amplitude does not yield information supporting the inflationary origin
of fluctuations against other possible generation
mechanisms\footnote{For an elaboration on this point, see \cite{holy}.}.
A determination of the tilt, on the other hand, would provide new and
powerful information. If observations were to find a small blue tilt,
this would rule out the standard inflationary paradigm, and it would
confirm a prediction first made in the context of superstring theory.
Note that the direct comparison between primordial fluctuations and
$B$-mode polarization is further complicated by the fact that there are
other sources of $B$-mode polarization, even on cosmological scales.
Quite generically, models with vector mode fluctuations also induce
$B$-mode polarization \cite{Durrer}. For example, cosmic strings can
produce direct $B$-mode polarization \cite{Gil}.

\section{Challenges for Bouncing Cosmologies}

\subsection{Addressing the Problems of Standard Big Bang Cosmology}

Inflationary cosmology is a successful solution to several problems
of Standard Big Bang cosmology, most importantly the {\it horizon, flatness,
entropy} and {\it structure formation} problems. We have already addressed
how the various bouncing models considered in this review can provide
a structure formation scenario which reproduces the successes of
inflation. But how are the other problems addressed \cite{Peter:2002cn,DBPP}?

All bouncing models discussed here solve the {\it horizon problem}. Since
time runs from $- \infty$, the horizon can be infinite and there is no
causal obstacle to explaining the isotropy of the cosmic microwave
background (the horizon problem).

Let us now turn to the {\it flatness problem}. In the case of inflationary
cosmology it is the accelerated expansion of space which dilutes
spatial curvature compared to the dominant contribution to the energy
density. The matter bounce scenario is neutral concerning the
impact of the spatial curvature term: in the case of a symmetric
bounce it decreases during the period of contraction by the same
amount that it increases during the expansion phase. In the Pre-Big-Bang
and Ekpyrotic scenarios the situation is better: the contribution of
spatial curvature decreases faster during the period of contraction
than it increases during the expanding phase. String gas cosmology,
on the other hand, does not solve the flatness problem. 

Inflation also provides an elegant mechanism to produce the large
entropy which our current universe contains: the exponential expansion
of space leads to an exponential increase in the total energy density
of matter. This energy density is converted into radiation during the
reheating phase at the end of inflation, producing a large entropy.
The entropy problem is only a problem if we assume that the universe
begins small (Planck size). The initial starting point in the matter bounce,
Pre-Big-Bang and Ekpyrotic cosmologies are completely orthogonal:
here it is assumed that the Universe starts large and cold. There is
then no problem whatsoever in explaining the current entropy.
If string gas cosmology is viewed as part of a bouncing cosmology,
the entropy problem is once again absent.
  
In the same way that inflationary cosmology faces some conceptual
problems, the alternatives discussed here also face some difficulties
of their own. We present some, deemed relevant, below.

\subsection{Initial conditions}

In any cosmological model it is necessary to specify initial
conditions. In a successful early universe model the initial
conditions should not have to be finely tuned. Whether
inflation is successful in this respect is an issue under
debate (see e.g. \cite{Tanmay, Gibbons} for arguments
claiming that the initial conditions for inflation need to
be finely tuned). Indeed, it can be shown that the
slow-roll trajectory which yields inflation is not a local
attractor in phase space in small field inflation models
(models in which the inflationary trajectory takes place
over field intervals $\delta \varphi < m_\mathrm{Pl}$ \cite{Piran}.
On the other hand, for large field inflation models the
slow-roll trajectory is a local attractor \cite{Kung}, even
in the presence of metric fluctuations \cite{Feldman}. For
a recent review of this issue the reader is referred to
\cite{RHBinflICrev}. Let us recall heuristically how
this works: one starts at a density somewhat below the
Planck density with a large and
inhomogeneous universe, with all field modes with
energy densities smaller than the Planck scale excited. 
Space will expand, and modes with wavelengths smaller
than the Hubble scale will be redshfited while those
with wavelengths larger will remain. In regions where
the field values for these modes is strictly positive,
inflation will then commence (see \cite{East} for a recent
numerical analysis of the dynamics).

Models of bouncing cosmologies must impose initial conditions 
at a time as far remote in the past as possible, and this 
corresponds to a moment when the Universe was large 
and mostly empty. As shown earlier in Fig.~\ref{FigBounce},
all relevant scales are then inside the Hubble radius at the time
the initial conditions are set up. Given the cosmological
background, it is hence natural to assume quantum vacuum
initial conditions for the fluctuations. On the other hand
it is not so easy to justify the initial conditions for the background
(see e.g. \cite{Buonanno} for a discussion of this point
in the context of Pre-Big-Bang cosmology). 

In string gas cosmology, it is assumed that the universe
begins as a hot gas of strings in a quasi-static
universe. Hence, thermal initial conditions for the
fluctuations are natural. Once again, there is a question
as to what produces the initial conditions for the background.

\subsection{Initial Inhomogeneities}

A bouncing scenario works opposite to inflation in the 
sense that one assumes the Universe begins large and 
then contracts. It would seem that any pre-existing inhomogeneities 
in a contracting universe will rapidly
and automatically collapse, leading to a highly inhomogeneous state.
This however is not obviously true, as the contraction may be
sufficiently slow compared to the diffusion rate of the primordial
constituents. In Ref.~\cite{PPNPN}, conditions were given for an
initially large contracting dust-dominated universe satisfying the Weyl
curvature hypothesis, to wipe out any primordial inhomogeneity
and hence to yield a satisfactory initial state out of which one can
settle vacuum initial conditions leading to our universe. It
should however be mentioned that, in this framework, the
presence, in the contracting phase, of a cosmological constant such as
that observed today, could easily destabilize the perturbations
produced in such a universe \cite{Maier:2011yy}, thereby leading
to predictions in disagreement with the current data.

\subsection{Anisotropies}

The question of shear in a contracting phase followed by a bounce is
central to setting constraints on such models. Indeed, in the
inflationary context, or even in most expanding cosmological models, the
shear, behaving as $a^{-6}$, rapidly becomes negligible compared
to any other component when the scale factor grows, and the
Friedman-Lema\^{\i}tre approximation of Eq.~\eqref{FLRW} can be safely
used. The contracting epoch preceding the bouncing phase is the exact
opposite and thus may induce a problem \cite{BCP}.

To illustrate this point, let us consider a spatially flat case and,
instead of the homogeneous and isotropic Eq.~\eqref{FLRW}, assume
spacetime to be well-described by an anisotropic Bianchi I metric (still
homogeneous) 
\begin{equation} 
\dd s^2 = \dd t^2 - a^2(t) \sum_{i=x,y,z}
\ex^{2\theta_i(t)} \dd x_i^2, \label{BianchiI} 
\end{equation} 
where $\sum_i\theta_i =0$. The Einstein equations then read 
\begin{equation} 
H^2 = \frac{8\pi\GN}{3} \rho + \frac16 \sum_i \dot \theta_i^2 =
\frac{8\pi\GN}{3} \left( \rho + \rho_\theta\right) \label{HBI}
\end{equation} 
and 
\begin{equation} 
\dot H = -4\pi\GN \left( \rho + p
\right) -\!\!\!\!\!\!\!\!\! \underbrace{\frac12 \sum_i \dot \theta_i^2
}_{\displaystyle 4\pi\GN \left( \rho_\theta + p_\theta \right)}\! \! \! \!\! \!\! \! ,
\end{equation} 
leading naturally to 
\begin{equation} 
\rho_\theta =
p_\theta = \frac{\sum_i\dot\theta_i^2}{16\pi\GN} \ \ \ \ \Longrightarrow
\ \ \ \ \ \rho_\theta \propto a^{-6} \label{w1} 
\end{equation}
as announced. Eq.~\eqref{w1} implies that the shear component rapidly
becomes negligible in a matter or radiation dominated expanding Universe,
and of course even more so in the almost exponential case of inflation.
During a phase of contraction however, as the scale factor shrinks to
zero, the shear can quickly come to dominate over all other components,
effectively ruining the FLRW approximation with the risk of transforming
a regular bounce into a Kasner singularity through the Belinsky
Khalatnikov Lifshitz (BKL) instability \cite{BKL}. If the initial shear
stems from primordial quantum fluctuations of the vector perturbations in
a vacuum state, then the resulting anisotropies remain at the same level
as the scalar ones. Provided the latter are well-behaved, the former must
therefore also remain small. If an initial classical shear is present
however, it will subsequently grow uncontrolled, thereby threatening the
entire scenario.

The most natural way out of the shear problem consists in postulating an
ekpyrotic phase, as discussed above. Indeed, this phase includes a
component, the scalar field $\phi$ with potential \eqref{PotEkp}, whose
equation of state is very large, $w\gg 1$, thus providing a contribution
$\rho_\phi \propto a^{-3(1+w)}$ which overcomes that of any anisotropy
that may have been originally present, hence preserving the FLRW nature
of the bounce \cite{noBKL}. Thus, the Ekpyrotic scenario is completely
safe from the anisotropy problem, and the matter bounce scenario
can be made safe by adding an Ekpyrotic phase of contraction at
higher curvatures. Some matter bounce models are in fact even
sensitive to the presence of radiation \cite{Karouby}.
See also \cite{otherAnis} for other approaches
to addressing the anisotropy problem.

\subsection{Relics}

The early universe, whatever the model (inflation or bounce), reaches
extremely high energy densities, close to the Grand Unified (GUT) or
even the string or Planck scales. This often provides an interesting
means of testing the otherwise unattainable theories supposed to be
valid at these scales, thus transforming cosmology into an invaluable
tool for high energy physics. Unfortunately, every coin has two faces,
and the very same attractive property implies a rather strong caveat in
the form of constraints: high energy theories usually predict loads of
new objects such as topological defects, exotic particle or even
primordial black holes (which we won't mention because estimates of
their remnant density differ by many orders of magnitude depending on
the model considered), which can be copiously produced during the early
stages of the universe, spoiling the subsequent evolution;
Ref.~\cite{Battefeld:2009sb} contains a short review of these relics.

The best known example stems from supersymmetric theories, and it is the
gravitino, i.e. the supersymmetric partner of the graviton. Depending on
the values of the coupling constants of the original theory, it can be a
stable particle, which may even be useful in cosmology in the form of
the missing dark matter. It can also be produced thermally at
high enough temperatures in the radiation dominated epoch. It is then
an example of a thermal relic. Stable relics may overclose the universe.
Unstable relics, on the other hand, could interfere with the process
of cosmic nucleosynthesis. 

The production of thermal relics depends on the final number density of
relics, and hence their contribution as dark matter, depends on particle
physics parameters and on the details of the cosmological evolution
during the early phases. If the scale of supersymmetry breaking is higher
than the maximal temperature in the expanding Big Bang phase (which in
the context of an inflationary model is the reheating temperature
$T_\mathrm{reheat}$) then gravitinos will not be produced and there are
no constraints. On the other hand, for a scale of supersymmetry breaking
which would explain the particle physics hierarchy problem and should be
not too much higher than $1 {\rm TeV}$, the relic particles will be
produced. Ref. \cite{Battefeld:2009sb} quotes the constraint
$T_\mathrm{reheat}\lesssim 4 \times 10^{10} {\rm Gev}$ for a gravitino
mass of about $1 {\rm TeV}$ which has to be satisfied in order that the
relics do not overclose the Universe.

Indeed, SUSY is a key ingredient in either GUT models or string theory,
and the natural energy scale associated with the corresponding theories
is expected to be much higher than the abovementioned value, and hence
the relic problem is an important one. However, the problem affects
bouncing cosmologies and simple inflationary models equally.

Topological defects in general are another very common prediction of 
particle physics models beyond the Standard Model. In particular,
in the context of a GUT theory with a high energy symmetry group 
$G$ which is simply connected, the vacuum manifold, i.e. the set
of field values which minimize the potential energy function
after the symmetry has broken to the Standard Model, is 
\begin{equation}
\mathcal{M} \sim G/[SU_\mathrm{c}(3)\times
U_\mathrm{em}(1)] \, , 
\end{equation}
and its second homotopy group
\begin{equation}
\pi_2 (\mathcal{M})\sim
\pi_1[SU_\mathrm{c}(3)\times U_\mathrm{em}(1)] \sim \pi_1[U(1)]\sim
\pi_1(S^1) \sim \mathbb{Z}
\end{equation}
is non trivial. This implies that pointlike topological defects,
monopoles, must be produced at some stage of the symmetry breaking
scheme. Their production in standard cosmology can be estimated by phase
transition arguments, and their number density, given their subsequent
evolution, is found much in excess of the closure density, thereby
ruining the entire universe's evolution. This problem was part of the
reason why inflation was proposed, as a such a phase naturally dilutes
the monopole number density in an exponential way: it suffices to ensure,
in a GUT implementation of inflation, that the monopole producing phase
takes place just before inflation, so that the monopoles are almost
instantaneously washed away.

In a bouncing framework however, monopoles will be
produced if the maximal temperature is higher than the
temperature of the monopole-forming phase transition (in the
same way that they are produced in inflation if the
reheating temperature is higher than this scale). An easy
way for bouncing cosmologies to avoid a potential
monopole problem is therefore that the maximal temperature
is less than the critical temperature at which symmetry
breaking takes place. Then, the symmetry whose breaking produces the
dangerous topological relics was thus never restored, and therefore
never broken. 

Another way to avoid a monopole problem which works independently
of the cosmological scenario is to start with a particle physics model
which does not at higher energies have a simply connected symmetry group.
Whereas this approach goes against the spirit of Grand Unification of the
1970s, it is quite realistic in the context of current string-based particle
physics models \cite{noGUT}.


Note that a bounce may also leave behind signals of the contracting phase
via a different dark matter distribution which is induced during contraction
\cite{Edna2}. 

\subsection{Instabilities}

A phase of contraction is subject to many more constraints than a phase
of expansion, because many new instabilities can take place. In particular,
scalar (curvature) and vector (shear) perturbations can grow too large
at the bounce transition. Ref.~\cite{DBPP} expands on these questions, which
we briefly consider below.

\subsubsection{Curvature}

In many bouncing models, the Bardeen potential develops a constant mode,
as in the usual inflationary scenario, but, in contrast to the latter,
the second mode is growing, and could pose a threat to the overall
treatment of the bounce as a background $+$ perturbations system. In
many situations, it was shown that this growing mode can remain under
control: there exists a set of conditions for the perturbative series to
make sense, and one can almost always find a gauge in which these
conditions are satisfied \cite{Vitenti:2011yc}. This also shows that not
all gauges represent valid descriptions of cosmological perturbations
near the bounce point, as the gauge-fixing conditions can become
undefined \footnote{Note that a similar problem arises during reheating in
inflationary cosmology.}.

Even if the possibly large amplitude of perturbation modes can be
tamed down to acceptable levels, their dependence in wavelength can
turn an almost scale invariant spectrum into a blue one, thus 
spoiling the predictions of an otherwise working model. Let us
illustrate this phenomenon with a phase of ekpyrotic contraction
leading to the bounce. The time evolution of a mode $\zeta_k$
corresponding to wavenumber $k$ of the comoving curvature
perturbation \eqref{zeta}  obeys
\begin{equation}
\zeta_k'' +2 \frac{z'}{z} \zeta_k'+c_\mathrm{S}^2 k^2 \zeta_k = 0,
\label{zeta_k}
\end{equation}
where the sound velocity $c_\mathrm{S}$ is given in terms of the pressure $p(X)$
[with $X\equiv \frac12 \left( \partial\varphi\right)^2$]
\begin{equation}
c_\mathrm{S}^2 = \frac{p_{,X}}{p_{,X} + 2 X p_{,XX}}
\label{csX}
\end{equation}
and $z\equiv a \sqrt{\dot H /(c_\mathrm{S} H)^2}$. The Mukhanov-Sasaki
mode variable $v_k = z \zeta_k$ satisfies 
\begin{equation}
v_k'' + \left( c_\mathrm{S}^2 k_2 -\frac{z''}{z}\right) = 0, 
\label{vk}
\end{equation}
which is nothing but the generalization, for a possibly time-varying sound
speed, of Eq.~\eqref{FourierEoM}.

Two problems may then arise. The first concerns the actual spectrum: in
the ekpyrotic contraction regime $c_\mathrm{S}^2\simeq 1$, and for long
wavelengths $k^2 \gg z''/z$, the solution reads $\zeta_k \sim
k^{-1/2}+\sqrt{k}\int z^{-2} \dd \eta +\cdots$, where vacuum initial
conditions have been taken into account. The constant solution term is
usually that which one considers as eventually producing the spectrum,
while the second term, much bluer, is always assumed negligible.
However, as shown in \cite{Xue:2011nw}, using the background equation of
motion $\dot H = -X p_{,X}$ and the definition of $z$, the contribution
due to the integral term can be estimated. The ratio between this
contribution and the supposedly dominant isocurvature component then
reads 
\begin{equation}
\frac{\zeta_k^\mathrm{int}}{\zeta_k^\mathrm{iso}}
\propto \ex^{\mathcal{N}_\mathrm{ekp} -2\mathcal{N}_k},
\label{ratio}
\end{equation}
with $\mathcal{N}_k$ the number of e-folds of ekpyrosis after the mode
$v_k$ has passed the potential $z''/z$ and $\mathcal{N}_\mathrm{ekp}$
the total duration, in e-folds, of the ekpyrotic phase. For modes of
cosmological relevance, one has $\mathcal{N}_k \simeq 10$, while
the constraint $\mathcal{N}_\mathrm{ekp}\gtrsim 60$, so \eqref{ratio}
spoils the overall mechanism, yielding a blue spectrum.

The second problem potentially induced by the evolution of curvature
perturbation concerns the bounce itself. Indeed, in order that it
actually takes place, a ghost condensate must be used, and therefore,
there must exist a finite amount of time around the bounce for which
$c_\mathrm{S}^2 \leq 0$. Now if, and this is a very model-dependent if,
during that time, there are modes satisfying $|c_\mathrm{S}^2 k^2 | >
z''/z$, then the solution naturally acquires an exponentially growing
term. Bounces such as that can be saved if either the bounce duration in
conformal time is sufficiently short, or if $|c_\mathrm{S}^2|$ is
sufficiently small (in practice exponentially small) during the
transition, so that the resulting exponential growth is not too large.

\subsubsection{Vector modes}

Although we did not consider vector modes in this short review, we must
mention them in the context of potential problems, as alluded to above.
In the usual paradigm of an expanding universe with ordinary matter
having no anisotropic stress (perfect fluid or scalar field in
practice), the vector modes are not sourced dynamically, and hence are
merely constrained to scale as $a^{-2}$. 

With the vector metric perturbation scaling as the inverse square of the
scale factor, one finds that for a fluid with constant equation of state $w$,
the velocity perturbation reads $V^i \propto k^2 a^{3w-1}$, which is
constant for a radiation dominated universe, and decreasing (resp. growing)
for any fluid having $w<\frac13$ (resp. $w>\frac13$) in an
expanding universe.

Assuming they were initially not dominant, this diluting factor renders
vector perturbations mostly harmless for standard cosmology and totally
irrelevant after even a short period of inflation took place. Obviously,
in a contracting universe, the situation can change drastically
\cite{Thorsten}. In the Pre Big-Bang scenario for instance, the
breakdown of perturbation theory seems unavoidable. It was however
argued in \cite{Pinto-Neto:2013zya} that if a primordial component of
vector perturbation were sourced by some dynamical vector field
beginning with vacuum initial conditions, their subsequent evolution,
although growing, should remain comparable to that of scalar
perturbations, and thus should not spoil the perturbation expansion.

It was also proposed \cite{Thorsten} to reverse this potential
catastrophe into a window of opportunity: assuming the growth to be
somehow controlled by, say, non linear effects, during the bouncing
phase, one could use the resulting relatively large amplitude vector
contribution to generate a large enough primordial magnetic field that
would explain the otherwise mysterious value necessary to understand
current data \cite{Kunze:2013kza}.

\subsection{Curvature and the Null Energy Condition}

Many bouncing models assume, to begin with, that the curvature is to be
neglected at all times, including at the bounce itself. The argument for
assuming flat spatial sections in a bouncing setup is that the curvature
term in the Friedman equation, $\Ka/a^2$, where $\Ka$ is the curvature
constant, will be effectively negligible
with respect to any other constituent that will necessarily be present
in the model, such as matter ($\rho_\mathrm{mat}\propto a^{-3}$) or
radiation ($\rho_\mathrm{rad}\propto a^{-4}$), in the limit where the
scale factor shrinks indefinitely $a\to 0$. In many cases of interest,
this argument is valid, but it may not be as generic as one would
spontaneously think.

First of all, in the context of general relativity, a regular bounce
with flat spatial sections can only take place provided the Null Energy
Condition $\rho+p\geq 0$ is violated. This is the main reason for
implementing bounces by means of negative energy scalar fields, ghost
condensates \cite{GhostCond}, conformal galileon \cite{ConfGal} and the
like \cite{ImplNEC}, sometimes leading to instabilities that have to
be dealt with \cite{NECviol}.

At the bouncing point, the Hubble length diverges, $H\to 0$, implying
that the curvature contribution, if any, ought to be exactly canceled by
the sum of all positive and negative energy contributions. Even though
this requirement might sound like fine-tuning, it actually is not: the
constraint then provides an algebraic equation giving the value of the
scale factor, $a_\mathrm{B}$ say, at which the bounce takes place, as a
function of the relative proportions of the various components involved
as well as on the curvature radius at that moment.

Another challenge induced by curvature is related to the question of
shear. Indeed, although the ekpyrotic phase permits to avoid the BKL
instability in the flat case \cite{noBKL}, the presence of curvature
could spoil the picture through a mixmaster phenomenon. Space would
expand and contract in different directions, leading to a highly
inhomogeneous and anisotropic universe \cite{CurvMix}.

\section{Conclusions}

Bouncing cosmologies still have a long way to go before they can be
considered as sound as the inflationary paradigm. Whereas
inflationary models are self-consistent at the level of an effective
field theory coupled to Einstein gravity, the same is not true of
bouncing models. On the other hand, the aim of bouncing cosmologies
is more ambitious in that one of their goals is to address the singularity
problem of our current cosmological models. To do this, one has to
go beyond a theory which is based on General Relativity coupled
to particle matter which obeys the usual energy conditions.
Another point to keep in mind is that inflationary models which look
self-consistent from the point of view of effective field theory may not
be consistent from the point of view of the complete ultraviolet
theory (see e.g. \cite{swampland} for a discussion of this problem).

Having built in a resolution of the primordial
singularity, a contracting phase followed by a bounce provides a natural
extension of the usual standard model of cosmology. There are
various scenarios in which a scale-invariant spectrum of cosmological
perturbations emerges which can explain all of the current data. We
have discussed the {\it matter bounce} model, the {\it Pre-Big-Bang}
and {\it Ekpyrotic} scenarios, and {\it string gas cosmology}. 

Theoretically, implementing a bouncing phase after a contraction era is
not so simple in the context of general relativity, as it entails a
violation of the Null Energy Condition (except in the unlikely event
that the spatial curvature somehow plays a crucial role). This requires
unusual scalar fields like a ghost condensate, with possibly many
resulting instabilities. Other implementations of a bouncing
scenario, like Pre-Big-Bang cosmology or the Ekpyrotic scenario,
invoke stringy effects to resolve the singularity and to yield a
bounce. String gas cosmology is based on fundamental principles
of superstring theory, but at the moment has no good implementation
in terms of an effective action. Thus, none of the bouncing
cosmologies considered here are at the present time
fully understood. Some may argue that inflation, demanding merely a
simple scalar field with a potential having a plateau \cite{Martin:2013nzq},
is perhaps simpler to achieve, and thus privileged from the Occam's
razor point of view. On the other hand, this philosophical standpoint,
useful as it is for providing guidance to write down underlying
theories, may not be valid for actual physical phenomena such as the
evolution of the entire universe.

From the point of view of cosmological observations, there is for the 
moment no need to go beyond inflationary cosmology: simple
single scalar field models are consistent will all cosmological observations.
On the other hand, many of the successful predictions of inflation
depend only on having a mechanism which produces an almost
scale-invariant spectrum of curvature fluctuations on scales which
are super-Hubble at early times, and thus the current observations
cannot be interpreted as favoring inflation. We have shown that
even the slight red tilt of the spectrum of curvature fluctuations 
(excluding purely scale-invariant perturbations at the
5$\sigma$ level) is not a unique prediction of inflation, but is naturally 
obtained in several bouncing models, in particular in string gas cosmology
and in the matter bounce. Similarly, the absence of measurable amounts 
of either non-Gaussianities or tensor modes is also obtained
naturally in string gas cosmology, although a simple matter
bounce is in tension with the data, whereas the data are in 
perfect agreement with inflationary predictions of simple inflation
models. 

We have discussed ways in which to distinguish between pure inflation
and pure bouncing cosmologies using future observational results. 
Of particular importance is the measurement of the amplitude and
slope of the tensor modes, since this will allow us to distinguish between
the consistency relations amongst observables predicted by
simple inflationary models on the one hand and bouncing cosmologies
(string gas cosmology in particular) on the other.

Of course, nothing prevents that the actual history of
our universe contains both an inflationary phase and a preceding bounce.
At least at the level of scalar field toy models for matter, most mechanisms 
for constructing a cosmological bounce allow
for the inclusion of an inflationary phase after the bounce.
At the time of writing, it is not clear if, in such a scenario, one will
ever be able to find truly discriminating measurements.

In this review we have studied bouncing cosmologies without ever
mentioning cyclic models. As we have seen, in the bouncing
models we have studied, a vacuum spectrum is transformed into
a scale-invariant one on scales which exit the Hubble radius during
the contracting period. In a model which is cyclic from the four space-time
dimensional point of view, the initial spectrum at the beginning of 
the second phase of contraction would be scale-invariant, and it
would be transformed into a spectrum with index $n_\mathrm{s} = -1$ before
the second bounce. This processing of cosmological fluctuations 
\cite{processing} makes four-dimensional cyclic cosmologies unpredictive.
The cyclic Ekpyrotic scenarios \cite{cyclic} avoid this problem since
it is new scales which are probed in each cycle (the model is not strictly
cyclic from the four-dimensional point of view). 

\section*{Acknowledgment} \noindent

One of the authors (RB) wishes to thank the Institute for Theoretical
Studies of the ETH Z\"urich for kind hospitality. He acknowledges
financial support from Dr. Max R\"ossler, the Walter Haefner Foundation
and the ETH Zurich Foundation, and from a Simons Foundation fellowship.
The research of RB is also supported in part by funds from NSERC and the
Canada Research Chair program. PP would like to thank the
Labex Institut Lagrange de Paris (reference ANR-10-LABX-63) part of the
Idex SUPER, within which this work has been partly done.

\end{document}